%% file: main.tex
\documentclass[sigconf]{acmart}

\AtBeginDocument{%
  }

\usepackage{placeins}
\usepackage{amsmath,amsfonts}
\usepackage{algorithm}
\usepackage{algpseudocode}
\usepackage{array}
\usepackage[utf8]{inputenc}
\usepackage[T1]{fontenc}
\usepackage{caption}
\usepackage{seqsplit}
\usepackage{longtable}
\usepackage{placeins}
\usepackage{makecell}
\usepackage{tcolorbox}
\usepackage{soul}
\usepackage{multirow}
\usepackage{adjustbox}
\usepackage{url}
\usepackage{verbatim}
\usepackage{graphicx}
\usepackage{booktabs}
\usepackage{xurl}
\usepackage{xcolor}
\usepackage{tabularx}

\setcopyright{acmlicensed} 
\copyrightyear{2026} 
\acmYear{2026} 
\acmDOI{XXXXXXX.XXXXXXX} 
\acmConference[CCS '26]{ACM Conference on Computer and Communications Security}{November 15-19, 2026}{The Hague, The Netherlands}  
\acmISBN{978-1-4503-XXXX-X/2018/06}  

\usepackage{tikz}
\usetikzlibrary{positioning,arrows.meta}
\usepackage{pgfplots}
\pgfplotsset{compat=1.18}
\usepgfplotslibrary{groupplots}
\usepackage{graphicx}
\usepackage{booktabs}
\usepackage{tabularx}
\usepackage{longtable}
\usepackage{multirow}
\usepackage{makecell}
\usepackage{array}
\usepackage{adjustbox}

\usepackage{placeins}
\usepackage{multicol}
\usepackage{enumitem}
\usepackage{ragged2e}
\usepackage{caption}
\usepackage{xurl}
\usepackage{listings}

\usepackage{amsmath,amsfonts}
\usepackage{algorithm}
\usepackage{algpseudocode}

\usepackage{pifont}
\newcommand{\cmark}{\ding{51}}
\usepackage{tcolorbox}
\usepackage{soul}
\usepackage{verbatim}

\newcolumntype{C}[1]{>{\centering\arraybackslash}m{#1}}
\newcolumntype{Y}{>{\RaggedRight\arraybackslash}X}

\DeclareRobustCommand{\findingbox}[2][gray!20]{%
\begin{tcolorbox}[
        left=0pt,
        right=0pt,
        top=0pt,
        bottom=0pt,
        colback=#1,
        colframe=#1,
        width=\linewidth, 
        enlarge left by=0mm,
        boxsep=5pt,
        arc=4pt,outer arc=4pt,  
        ]
        #2
\end{tcolorbox}
}

\begin{document}

\title{An Empirical Study of Observability Limits in Advanced Software Supply Chain Attacks}

\settopmatter{authorsperrow=3}

\author{Zhuoran Tan}
\authornote{Both authors contributed equally to this work.}
\affiliation{%
  \institution{University of Glasgow}
  \city{Glasgow}
  \country{United Kingdom} 
}
\email{z.tan.1@research.gla.ac.uk}

\author{Wenbo Guo}
\authornotemark[1]
\affiliation{%
  \institution{Nanyang Technological University}
  \city{Singapore}
  \country{Singapore}
}
\email{honywenair@gmail.com}

\author{Jiewen Luo}
\affiliation{%
  \institution{Royal Holloway, University of London}
  \city{Egham}
  \country{United Kingdom}
}
\email{Jiewen.LUO.2016@live.rhul.ac.uk}

\author{Taylor Brierley}
\affiliation{%
  \institution{JUMPSEC Ltd}
  \city{London}
  \country{United Kingdom}
}
\email{taylorbrierley03@gmail.com}

\author{Jeremy Singer}
\affiliation{%
  \institution{University of Glasgow}
  \city{Glasgow}
  \country{United Kingdom}
}
\email{Jeremy.Singer@glasgow.ac.uk}

\author{Christos Anagnostopoulos}
\affiliation{%
  \institution{University of Glasgow}
  \city{Glasgow}
  \country{United Kingdom}
}
\email{Christos.Anagnostopoulos@glasgow.ac.uk}

\renewcommand{\shortauthors}{Tan et al.}

\input{sections/abstract.tex}



\maketitle

\input{sections/introduction.tex}

\input{sections/motivation.tex}

\input{sections/methodology.tex}

\input{sections/analysis.tex}

\input{sections/discussion.tex}

\input{sections/conclusion.tex}

\begin{acks}
We thank Marc Juarez and Sebastian Garcia for their helpful
discussions and suggestions. We also thank Max Corbridge,
Chris Preece, and Matt Lawrence from JUMPSEC Ltd for their
early feedback and technical guidance, and Scott MacKintosh
for his support with the Azure infrastructure.
\end{acks}

\appendix

\section*{Open Science}
We will release the following public artifacts to enable evaluation and reproduction of all experiments:
\begin{itemize}
  \item \textbf{SynthChain dataset (sanitized)}: system telemetry, provenance information, and ground-truth annotations for all scenarios.
  \item \textbf{Experiment code}: preprocessing, feature extraction, training/evaluation scripts, and configuration files.
  \item \textbf{Baselines and instructions}: implementation details and step-by-step commands to reproduce each table/figure.
  \item \textbf{Documentation}: environment setup, dependencies, and a reproducibility checklist.
\end{itemize}
All public artifacts are now available at: \url{https://anonymous.4open.science/r/SSCMDataset-2E11}.

\section*{Ethical Considerations}
\label{app:ethical_considerations}
SynthChain is a benchmark dataset providing multi-source runtime telemetry with chain-level ground truth for evaluating forensic reconstruction and detection of SSC attacks.
The main ethical risk is misuse of realistic attack simulation and payload orchestration.
To mitigate this risk, we publicly release only the artifacts required to evaluate our claims
(system telemetry, provenance, and ground-truth annotations), and we do \emph{not} publicly
release executable payloads or end-to-end attack orchestration.

During peer review, we provide an anonymized evaluation artifact to reviewers to support
reproducibility. After acceptance, we will release a sanitized version of the dataset and
experiment code, with any payloads/malware removed. Access to the full attack simulation
code (payload implementations and orchestration) is provided under controlled access only,
to verified researchers for academic or defensive purposes, following responsible disclosure
practices.

\bibliographystyle{ACM-Reference-Format}
\bibliography{references}

\section{Benign Activity Simulation}
\label{app:benign-activity}

To evaluate detection and forensic analysis under realistic runtime conditions, we introduce a benign background activity simulation framework that generates diverse, non-deterministic system behaviors concurrent with attack execution. The goal of this framework is not to precisely emulate human intent or productivity workflows, but to introduce representative operational noise commonly observed on real-world developer and end-user systems.

\subsection{Design Objectives}

The benign activity is designed to satisfy three objectives:
\begin{enumerate}
    \item \textbf{Behavioral Coverage}: generate system- and network-level events that overlap with those produced by early-stage supply chain attacks (e.g., process creation, file I/O, outbound connections).
    \item \textbf{Temporal Variability}: avoid deterministic execution patterns by randomizing activity timing and selection.
    \item \textbf{Role Diversity}: reflect heterogeneity across hosts by assigning different functional profiles.
\end{enumerate}

\begin{table*}[t]
\centering
\caption{Behaviour definition (keywords): normal vs. malicious activities instantiated in our scenarios.}
\label{tab:behaviour_definition}
\small
\setlength{\tabcolsep}{4pt}
\renewcommand{\arraystretch}{1.02}
\begin{tabular}{@{}l p{6.0cm} p{6.0cm}@{}}
\toprule
\textbf{Behaviour Category} & \textbf{Normal (keywords)} & \textbf{Malicious (keywords)} \\
\midrule
Process \& Script Execution &
Browser/editor/terminal; tests; benign scripts &
Install-hook exec; staged scripts; PowerShell/assembly; interpreter spawning \\
\hline
Package / Dependency Ops &
\texttt{pip}/\texttt{npm} install/update; repo clone; dependency fetch &
Typosquat packages; staged deps; install$\rightarrow$exec; multi-stage deps \\
\hline
External Communications &
Browsing/search; legit APIs; SSH to known hosts; updates &
C2 callbacks; module pull; long-lived encrypted channels; exfil to attacker \\
\hline
Discovery \& Collection &
Code/file edit; test artifacts; routine file copy (SCP) &
Host/file enumeration; writable-path search; archive (\texttt{*.zip}); staging \\
\hline
Credentials \& Privilege &
Normal logins; user sessions; routine secret use &
Secret access; token abuse; privilege escalation; credential dumping attempts \\
\hline
Services \& Listening &
Expected dev services (web/DNS/DB/SSHD); routine ports &
Adversary listeners; anomalous binds; temporary service for C2/exfil \\
\hline
Container / CI/CD &
Build images; run containers; deploy actions; pipeline ops &
Pipeline credential abuse; artifact tamper; build-stage exec; CI/CD retrieval \\
\bottomrule
\end{tabular}
\end{table*}

\subsection{Activity Categories}
Each simulated host executes a subset of benign activities drawn from the following categories:
\begin{itemize}
    \item \textbf{Web and Network Interaction}: outbound web browsing, search queries, and API-based communications.
    \item \textbf{Remote Administration}: authenticated remote access and command execution.
    \item \textbf{File Operations}: file copying, directory creation, and document handling.
    \item \textbf{System Maintenance}: periodic software updates and package management.
    \item \textbf{Development Workflows}: source code retrieval, test execution, and application deployment.
\end{itemize}

These activities collectively produce realistic background signals, including filesystem modifications, process lifecycles, network flows, and authentication traces, which may partially overlap with attack-related telemetry.

\subsection{Scheduling and Execution Logic}
Benign activities are scheduled probabilistically during predefined working hours. At each scheduling opportunity, a single activity is randomly selected and executed, ensuring the benign workloads interleave with attack behaviours rather than being isolated in separate execution windows. Algorithm~\ref{alg:normal-activity} summarizes the high-level scheduling logic:

\begin{algorithm}[t]
\caption{Normal Activity Simulation}
\label{alg:normal-activity}
\begin{algorithmic}[1]
\Require
ActivitySet = \{Web, RemoteAccess, FileOp, Update, Download, Dev, API, Login\} \\
ActiveHours = [09{:}00, 19{:}00] \\
$N$ = number of scheduled activities
\For{$i = 1$ to $N$}
    \State $delay \gets \text{RandomInterval}(min, max)$
    \State \Call{Wait}{$delay$}
    \If{\text{CurrentTime} $\in$ ActiveHours}
        \State $activity \gets \text{RandomChoice}(\text{ActivitySet})$
        \State \Call{Execute}{$activity$}
    \EndIf
\EndFor
\end{algorithmic}
\end{algorithm}

Each activity invocation may trigger multiple low-level events (e.g., child processes, network connections, or file writes), thereby generating compound benign traces rather than isolated actions.

\subsection{Scope and Limitations}
The benign activity simulation is intentionally lightweight and abstract. We do not attempt to model fine-grained human intent, productivity cycles, or organizational policies. Instead, the framework provides sufficient benign variability, as shown in Table~\ref{tab:behaviour_definition}, to challenge detection and forensic analysis while maintaining reproducibility and experimental control.

\section{Extended Statistical Analysis of Malicious Packages}
\label{app:malicious package statistic analysis table}

To characterize current exploitation trends in SSC attacks, we analyze the OpenSSF dataset collected through 2025~\cite{malicious-packages}, which contains 16,272 malicious packages across four major ecosystems (npm, PyPI, RubyGems, Rust) along with metadata describing their malicious behaviors.

For each package, we extract six dimensions capturing both structural and behavioural characteristics from description of individual packages: Ecosystem (platform), Location (where malicious code is embedded), Function (intended malicious action), Attack Type (high-level exploitation pattern), Trigger mechanism (conditions activating the behaviour), and Evasion method (techniques used to avoid detection).

Figure \ref{fig:pkg-taxonomy} summarises the aggregated distributions of these behaviours. It shows a highly skewed distribution across several dimensions: entrypoint/download and setup-script injection account for roughly 56.4\% and 33.6\% of all packages, respectively ($\approx$90\% combined), mirroring the dominance of install-time behaviors in both function and attack-type labels. Triggering is overwhelmingly installation-centric ($\approx$95.8\% of packages are activated upon installation among those with a trigger label). Evasion labels are present in only a subset of packages ($\approx$39.6\%); within this subset, Base64 encoding constitutes $\approx$84.7\% of all evasion instances, indicating that lightweight transformation remains the primary stealth strategy, while techniques such as payload splitting or steganography appear only in the long tail.

Additionally, most malicious packages are triggered \emph{upon installation or download} (15,583 cases), confirming installation-time activation as the dominant entry point. For evasion, lightweight techniques are overwhelmingly prevalent: Base64-based encoding alone appears in over 5,000 packages, far exceeding more sophisticated approaches such as payload splitting or steganography.

These findings indicate three consistent regularities:
\begin{enumerate}
\item \textbf{Install-time execution is the primary activation strategy for initial foothold establishment};
\item \textbf{Payload delivery and data theft/exfiltration are the central objectives}; 
\item \textbf{Simple but effective evasion (especially encoding/ obfuscation) is favoured by attackers}.
\end{enumerate}

\begin{figure}[t]
\centering
\begin{tikzpicture}
\begin{groupplot}[
  group style={group size=2 by 1, horizontal sep=2.2cm},
  width=0.38\columnwidth,
  height=4.8cm,
  xbar,
  xmin=0,
  scaled x ticks=false,
  xmajorgrids,
  enlarge y limits=0.22,
  y dir=reverse,
  tick align=outside,
  x tick label style={
    font=\scriptsize,
    rotate=45,
    anchor=north east,
    /pgf/number format/fixed,
    /pgf/number format/precision=0,
    /pgf/number format/1000 sep={,},
  },
  xlabel style={font=\scriptsize},
  nodes near coords,
  nodes near coords align={horizontal},
  every node near coord/.append style={font=\scriptsize},
  every axis title/.append style={font=\small},
  ytick=data,
  yticklabel style={font=\scriptsize},
]
\nextgroupplot[
  title={Trigger Mechanisms},
  xlabel={Count},
  xmax=20000,
  xtick={0,5000,10000,15000,20000},
  symbolic y coords={
    {Upon Installation},
    {Network Activity},
    {Function Call},
    {Async Install},
    {Upon Usage},
    {Conditional Config}},
]
\addplot coordinates {
  (15583,{Upon Installation})
  (594,{Network Activity})
  (21,{Function Call})
  (14,{Async Install})
  (10,{Upon Usage})
  (1,{Conditional Config})
};
\nextgroupplot[
  title={Evasion Methods},
  xlabel={Count},
  xmax=7000,
  xtick={0,2000,4000,6000},
  symbolic y coords={
    {Base64 Encoding},
    {Obfuscation},
    {Split Payloads},
    {Steganography}},
]
\addplot coordinates {
  (5460,{Base64 Encoding})
  (900,{Obfuscation})
  (56,{Split Payloads})
  (27,{Steganography})
};
\end{groupplot}
\end{tikzpicture}
 
\vspace{6pt}
 
\small
\setlength{\tabcolsep}{3pt}
\begin{tabular}{@{}rlllr@{}}
\toprule
\textbf{\#} & \textbf{Trigger Location} & \textbf{Function} & \textbf{Objective} & \textbf{Count} \\
\midrule
1 & Entrypoint/Download & Install malware     & Data exfiltration  & 9{,}179 \\
2 & Setup script          & Spyware/info steal & Dropper  & 5{,}426 \\
3 & Variables             & Crypto miner         & Typosquatting      &   900 \\
4 & Comm. module  & Maintain C2          & C2 channel         &   594 \\
5 & Special functions     & Persistence          & Cmd execution      &    46 \\
6 & Hooked seq.\ files   & Steal secrets        & Starjacking        &    42 \\
7 & ---                   & Credential theft     & Social engineering &    41 \\
8 & ---                   & Parallel download & ---                &    14 \\
\bottomrule
\end{tabular}
 
\vspace{2pt}
\caption{Distributions of malicious semantics and trigger/evasion mechanisms.}
\label{fig:pkg-taxonomy}
\end{figure}

\section{Extended Case Study — SC1 (Stegano)}
\label{app:extended sc1}

This appendix provides the full telemetry breakdown and supporting evidence for the SC1 exemplar.
SC1 (Stegano) was executed on a single Windows 10 virtual machine to emulate a victim developer endpoint installing a typosquatted Python package. We monitored the system for a 189-minute observation window (12:28--15:37 UTC) and collected host and network telemetry using Azure Monitor Agent (AMA). AMA was configured to export six telemetry streams covering process execution, network connections, Windows security events, system events, bound ports, and performance counters. In total, the collection yields 23,534 records and provides complementary visibility into (i) process-level execution context, (ii) outbound connection behavior, and (iii) authentication/privilege-related security events (Table~\ref{tab:sc1_data_overview}).

\begin{table}[htbp]
\centering
\caption{SC1 data collection statistics across six telemetry sources.}
\label{tab:sc1_data_overview}
\small
\begin{tabular}{@{}lrr@{}}
\toprule
\textbf{Data Source} & \textbf{Records} & \textbf{Size} \\
\midrule
Process execution (VMProcess) & 345 & 328 KB \\
Network connections (VMConnection) & 5,746 & 3.4 MB \\
Security events (SecurityEvent) & 334 & 440 KB \\
System events (Event) & 1,000 & 2.3 MB \\
Bound ports (VMBoundPort) & 1,109 & 503 KB \\
Performance counters (Perf) & 15,000 & 5.5 MB \\
\bottomrule
\end{tabular}
\end{table}

\subsection{Experimental Setup and Data Sources}

This scenario captures a supply chain attack in which a typosquatted Python package delivers a steganographically-concealed command-and-control agent. The malicious package \texttt{colorsapi-6.6.7} mimics the legitimate \texttt{colorapi} library through single-character insertion. When victims install this package, the \texttt{setup.py} script executes pre-installation hooks that download an image file from a content delivery network. Hidden within this image is executable Python code, embedded using Least Significant Bit (LSB) steganography, which establishes a persistent backdoor on the victim system.

\subsection{Step-by-step Attack Timeline}

\textbf{Phase 1: Initial Access (13:22 UTC).} The victim runs \texttt{pip install colorsapi}, which triggers execution of the typosquatted package’s \texttt{setup.py} during installation.

\textbf{Phase 2: Payload Retrieval (13:22 UTC).} The installer downloads an 8.6\,MB PNG from a CDN endpoint (\texttt{146.75.74.132}) and extracts embedded code via LSB steganography (T1027.003).

\textbf{Phase 3: Code Execution (13:23 UTC).} The extracted Python payload is executed in-process (via \texttt{exec()}), spawning a Mythic C2 agent (Medusa variant) under the Python interpreter context.

\textbf{Phase 4: C2 Establishment (13:28 UTC).} The agent establishes an SSH channel to \texttt{172.187.202.111:22} and sustains activity for $\sim$120 minutes, transferring $\sim$265\,MB of modular Python tooling (loaded in-memory).

\textbf{Phase 5: Collection and Exfiltration (14:33--15:18 UTC).} The attacker runs a reconnaissance script (\texttt{SenScanner.py}), packages results into \texttt{info.zip}, and exfiltrates data over the existing SSH tunnel, completing at 15:18 UTC.

\subsection{Host-level Evidence: Process Telemetry}

Process telemetry reveals a clear separation between attack-related and benign activity. Of the 345 recorded processes, 73 (21.2\%) are Python interpreter instances, and the command \texttt{python3 setup.py install} appears four times throughout the observation period, each invocation spawning additional child processes through Python's multiprocessing module. Table~\ref{tab:sc1_process_breakdown} categorizes all observed processes by their behavioral role, distinguishing malicious execution from legitimate user activity and system services.

\begin{table}[htbp]
\centering
\caption{SC1 process categorization by behavioral role.}
\label{tab:sc1_process_breakdown}
\small
\begin{tabular}{@{}llr@{}}
\toprule
\textbf{Category} & \textbf{Representative Processes} & \textbf{Count} \\
\midrule
Malicious execution & python, python3 & 73 \\
User activity & chrome, msedge, Explorer & 48 \\
System services & svchost, HealthService & 156 \\
Development tools & git-remote-https & 3 \\
Other & SearchApp, OneDrive & 65 \\
\bottomrule
\end{tabular}
\end{table}

\subsection{Network-level Evidence: Infrastructure and Traffic Patterns}

\begin{figure*}[t]
\centering
\includegraphics[width=0.85\textwidth]{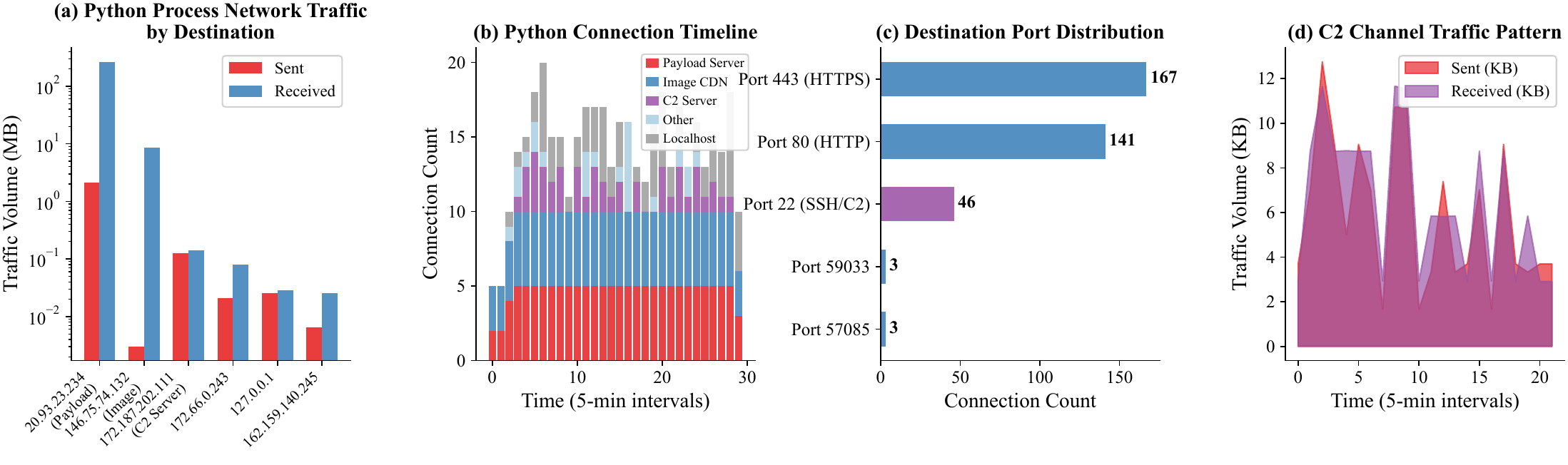}
\caption{SC1 network traffic analysis. (a) Traffic volume per destination IP on logarithmic scale, showing 265 MB received from the payload server versus 2.1 MB sent. (b) Connection timeline in 5-minute intervals, with sustained C2 activity (purple) and payload retrieval (red) visible throughout the attack window. (c) Destination port distribution, where SSH (port 22) accounts for 46 connections exclusively from Python processes. (d) C2 channel traffic pattern exhibiting irregular beacon intervals ($\mu = 4.2$ min, $\sigma = 1.8$ min).}
\label{fig:sc1_network}
\end{figure*}

Network analysis exposes the attack infrastructure. Python processes established 427 connections to six distinct IP addresses, with traffic patterns that diverge sharply from normal development workflows. Table~\ref{tab:sc1_network_summary} breaks down these connections by destination. The C2 server at \texttt{172.187.202.111} received 46 SSH connections carrying 277 KB of bidirectional traffic. More notably, the payload server at \texttt{20.93.23.234} delivered 265.4 MB to the victim, while the image CDN at \texttt{146.75.74.132} transferred 8.6 MB containing the steganographic payload. Figure~\ref{fig:sc1_network}(a) visualizes this traffic asymmetry on a logarithmic scale, where received bytes exceed sent bytes by two orders of magnitude for the payload server.

\begin{table}[htbp]
\centering
\caption{SC1 Python process network communication by destination.}
\label{tab:sc1_network_summary}
\footnotesize
\begin{tabular}{@{}llrrr@{}}
\toprule
\textbf{Role} & \textbf{Destination IP} & \textbf{Port} & \textbf{Sent} & \textbf{Received} \\
\midrule
C2 Server & 172.187.202.111 & 22 & 128 KB & 143 KB \\
Payload Server & 20.93.23.234 & 80 & 2.1 MB & 265.4 MB \\
Image CDN & 146.75.74.132 & 443 & 3 KB & 8.6 MB \\
Package Index & 172.66.0.243 & 443 & 22 KB & 82 KB \\
Localhost (IPC) & 127.0.0.1 & varied & 26 KB & 29 KB \\
\bottomrule
\end{tabular}
\end{table}

\subsection{Security Event Context}

Windows Security logs recorded 334 events, dominated by Credential Manager reads (Event ID 5379, 91 occurrences). Successful logon events (ID 4624) and special privilege assignments (ID 4672) each appeared 69 times, while 10 failed logon attempts (ID 4625) were recorded, none of which correlated temporally with attack activity.

\subsection{Evasion and Operational Characteristics}

The attack employs multiple evasion techniques that compound detection difficulty. Steganographic delivery ensures the initial payload reaches the victim without triggering content-aware firewalls, as the carrier image passes standard file-type validation. The use of Fastly's CDN infrastructure provides additional cover; blocking this IP would disrupt access to thousands of legitimate websites. The SSH-based C2 channel exploits the protocol's ubiquity in development environments, where connections to unfamiliar servers may not raise immediate suspicion. Most critically, the 265 MB in-memory module loading enables fileless operation during the reconnaissance and exfiltration phases, leaving no on-disk artifacts for endpoint detection tools to discover.

Temporal analysis reveals operational security awareness. The 70-minute payload download (13:33 to 14:43) proceeds at an average rate of 63 KB/s, well below thresholds that might trigger bandwidth-based anomaly detection. Figure~\ref{fig:sc1_network}(d) shows that C2 beacon intervals follow a jittered pattern ($\mu = 4.2$ min, $\sigma = 1.8$ min) rather than fixed periodicity, frustrating detection rules based on regular callback timing. The total attack duration of 116 minutes exceeds the analysis window of most automated sandboxes, which typically terminate after 10 to 15 minutes of execution.

\findingbox{Finding SC1: Three behavioral anomalies distinguish this attack from legitimate activity. (1) \textbf{Traffic asymmetry}: Python processes received 274 MB while transmitting only 2.3 MB, yielding a 119:1 download-to-upload ratio inconsistent with typical API interactions or package installations. (2) \textbf{Process-protocol mismatch}: SSH connections originated from \texttt{python3.exe} rather than standard SSH clients such as \texttt{ssh.exe} or PuTTY, violating expected process-to-protocol mappings. (3) \textbf{Working directory correlation}: All 73 Python processes share the installation path \texttt{C:\textbackslash xx\textbackslash xx\textbackslash colorsapi-6.6.7\textbackslash} as their working directory, enabling attribution of the entire attack chain through a single forensic pivot point.}

\section{Metric Definitions}
\label{app:metrics}

Let $\mathcal{E}_s$ denote the set of expected coarse steps for scenario $s$, with $E_s = |\mathcal{E}_s|$. Let $\mathcal{T}_{s,c}$ denote the set of step types observed (tagged) under configuration $c$, and $\mathcal{C}_{s,c} \subseteq \mathcal{T}_{s,c}$ the subset that appears in the reconstructed chain.

\paragraph{Coverage.}
\begin{equation}
\mathrm{TagCov}(s,c) = \frac{|\mathcal{T}_{s,c}|}{E_s}, \qquad
\mathrm{ChainCov}(s,c) = \frac{|\mathcal{C}_{s,c}|}{E_s}.
\end{equation}

\paragraph{Precision and Recall.}
Step-level precision and recall measure how well the observed steps match the expected set:
\begin{equation}
\mathrm{StepP}(s,c) = \frac{|\mathcal{T}_{s,c} \cap \mathcal{E}_s|}{|\mathcal{T}_{s,c}|}, \qquad
\mathrm{StepR}(s,c) = \frac{|\mathcal{T}_{s,c} \cap \mathcal{E}_s|}{|\mathcal{E}_s|}.
\end{equation}
Chain-level variants replace $\mathcal{T}_{s,c}$ with $\mathcal{C}_{s,c}$:
\begin{equation}
\mathrm{ChainP}(s,c) = \frac{|\mathcal{C}_{s,c} \cap \mathcal{E}_s|}{|\mathcal{C}_{s,c}|}, \qquad
\mathrm{ChainR}(s,c) = \frac{|\mathcal{C}_{s,c} \cap \mathcal{E}_s|}{|\mathcal{E}_s|}.
\end{equation}
When $|\mathcal{T}_{s,c}| = 0$ (or $|\mathcal{C}_{s,c}| = 0$), precision is undefined; we conservatively set it to zero.

\paragraph{Continuity proxy.}
For each step $i$ in the reconstructed chain, let $[\min\_ts_i,\, \max\_ts_i]$ be its time window. For adjacent steps $(i, i{+}1)$, the inter-step gap is $\delta_i = \min\_ts_{i+1} - \max\_ts_i$. The continuity proxy is:
\begin{equation}
\mathrm{Cont}(s,c) = \frac{|\{i : \delta_i \leq W\}|}{|\text{adjacent pairs}|},
\end{equation}
where $W = 600$\,s by default.

\paragraph{Reconstructability.}
\begin{equation}
\mathrm{Recon}(s,c) = \mathrm{StepR}(s,c) \cdot \mathrm{Cont}(s,c).
\end{equation}

\paragraph{Cross-scenario aggregation.}
Coverage and recall are aggregated as weighted means (weight $w_s = E_s / \sum_{s'} E_{s'}$, $\sum_s w_s = 1$):
\begin{equation}
\overline{M}_{c}^{\,\mathrm{wtd}} = \sum_{s \in \mathcal{S}_c} w_s \cdot M(s,c).
\end{equation}
Precision and reconstructability use the unweighted mean:
\begin{equation}
\overline{M}_{c}^{\,\mathrm{mean}} = \frac{1}{|\mathcal{S}_c|} \sum_{s \in \mathcal{S}_c} M(s,c).
\end{equation}

\section{Case Studies for Remaining Scenarios (Brief Summaries)}
\label{app:remaining case studies}

\paragraph{SC2: Starter --- Persistence via Startup Folder}
\label{sec:sc2_starter}
SC2 models a lightweight supply-chain payload that establishes persistence by placing an autostart artifact in the user Startup folder, enabling execution on subsequent logins. In our telemetry, the most reliable anchors are host-side persistence signals (file/registry updates consistent with startup configuration) and coarse execution evidence around the initial drop. Under full telemetry, reconstruction reaches $\mathrm{StepR}=0.75$ by recovering \texttt{INSTALL}, \texttt{DOWNLOAD}, and \texttt{EXFIL}, but still misses a clean \texttt{\seqsplit{OUTBOUND\_CONN}} anchor, illustrating a common ``incomplete chain'' pattern where persistence is observable while network establishment lacks joinable attribution. Single-source settings further underperform because system logs alone cannot consistently connect startup persistence to subsequent network behavior without process-to-connection linkage.

\paragraph{SC3: Parallel --- Multi-Script Concurrent Execution}
\label{sec:sc3_parallel}

SC3 benefits from multi-source host+network telemetry, improving the best single-source recall (0.25) to 0.50 under full telemetry. Nevertheless, \texttt{DOWNLOAD} and \texttt{EXFIL} remain missing from the expected coarse-step set, suggesting that concurrent benign-like network activity and multi-process overlap reduce the distinctiveness of retrieval and exfiltration phases. The dominant observable anchors are \texttt{INSTALL} and \texttt{\seqsplit{OUTBOUND\_CONN}}, which can be correlated temporally but do not uniquely determine end-to-end intent without stronger content or file-transfer evidence.

\paragraph{SC5: 3CX --- Multi-Stage Backdoor Deployment}
\label{sec:sc5_3cx}
SC5 is constrained by evidence availability: the dataset is dominated by \texttt{events} (treated as a composite stream) and yields $\mathrm{StepR}=0.25$, observing primarily \texttt{INSTALL}-adjacent activity (plus extra \texttt{AUTH}) while missing \texttt{DOWNLOAD}, \texttt{\seqsplit{OUTBOUND\_CONN}}, and \texttt{EXFIL}. This scenario illustrates that even when an exported stream is internally multi-channel, it may still lack the specific join keys (e.g., process-to-network attribution) required to reconstruct later-stage communication.

\paragraph{SC6: CloudEX --- CI/CD Pipeline Compromise}
\label{sec:sc6_cloudex}
SC6 remains difficult under available telemetry: even with \texttt{syslog+events} the reconstruction achieves $\mathrm{StepR}=0.25$ and primarily tags \texttt{AUTH} (expected) plus an extra \texttt{INSTALL}, while missing \texttt{DOWNLOAD}, \texttt{\seqsplit{OUTBOUND\_CONN}}, and \texttt{EXFIL}. This is consistent with a control-plane dominated attack where critical actions occur in cloud identity/API layers that are not fully captured (or not captured in a rule-matchable schema) by the current logging exports, emphasizing the need for IAM/API audit streams \cite{287296} and better cloud-identity step rules.

\paragraph{SC7: LayerInj --- Neural Network Model Backdoor}
\label{sec:sc7_layerinj}
SC7 demonstrates a precision trade-off under multi-source correlation. The best single-source configuration (Suricata) already achieves $\mathrm{StepR}=0.667$ with perfect precision ($\mathrm{StepP}=1.0$) by capturing \texttt{\seqsplit{OUTBOUND\_CONN}} and \texttt{EXFIL}. Full telemetry maintains the same recall (0.667) but reduces precision to 0.5 by introducing extra step candidates (\texttt{AUTH}, \texttt{INSTALL}) not present in the expected coarse-step set, reflecting the risk of over-attributing generic system events to an attack narrative when the core maliciousness is semantic (model behavior) rather than OS-level execution novelty.

\section{Attack Flow \& Data Quality Discussion}
\label{app:attack-flows}

This section provides a structured overview of the simulated attack scenarios and their execution flows. 
We summarize the key characteristics of each scenario and present representative attack diagrams to illustrate how multi-stage behaviors are triggered and executed in practice. 
These scenarios are designed to cover diverse attack surfaces, including software supply chain compromise, script-based execution, and cloud-based intrusion, with varying triggering mechanisms and evasion strategies.

Importantly, the scenarios are constructed with explicit observability in mind: each attack step is associated with concrete system-level artifacts (e.g., network traffic, file operations), enabling IOC-based labeling. 
This design allows us to analyze not only the attack behaviors themselves but also the extent to which they can be captured through observable indicators. 
Accordingly, we further examine IOC coverage and data quality to characterize the completeness and limitations of the resulting ground truth.

\subsection{Attack Scenarios and Diagram}

Table~\ref{tab:attack scenarios} summarizes the seven simulated attack scenarios. Each scenario is characterized along several dimensions: (i) \textit{initial trigger}, (ii) \textit{attack steps}, describing the sequence of critical steps; (iii) \textit{evasion techniques}, such as obfuscation or staged execution; and (iv) \textit{attack objectives}, primarily focused on data exfiltration.

\begin{table*}[t]
    \centering
    \caption{Overview of Attack Scenarios}
    \label{tab:attack scenarios}
    \footnotesize
    \begin{tabular}{@{}lp{4.5cm}p{1.8cm}p{2.2cm}p{1.5cm}p{2cm}c@{}}
        \toprule
        \textbf{Case} & \textbf{Critical Attack Steps} & \textbf{Trigger} & \textbf{Evasion Techniques} & \textbf{Objectives} & \textbf{Tools} & \textbf{OS}\\
        \midrule
        Stegano & (1) Install typosquatting package; (2) Install image embedded with malware & setup.py & Obfuscation, steganography, erase trace & Exfiltrate system data (C2) & Medusa~\footnote{https://github.com/MythicAgents/Medusa} & WIN\\[1ex]
        \hline
        Starter & (1) Install typosquatting package; (2) Create startup folder; (3) Local file replacement; (4) Install exe scripts & setup.py & Stream cipher, file replacement, erase trace & Exfiltrate system data (C2) & Mimikatz~\footnote{https://www.kali.org/tools/mimikatz/}, Powershell, Apollo~\footnote{https://github.com/MythicAgents/Apollo}  & WIN\\[1ex]
        \hline
        Parallel & (1) Install malicious package; (2) package.json triggers preinstall.js; (3) Initiate index.js; (4) Compress scanned info & Inter-hooked scripts & Separate running & Exfiltrate system (HTTP) / sensitive info (FTP) & --- & Linux\\[1ex]
        \hline
        NPMEX & (1) Outbound connection triggered upon package download;
        (2) Install two malicious NPM dependencies sequentially;
        (3) 1st package retrieves token from remote server;
        (4) 2nd package fetches and executes staged payload &
        Upon download &
        Run in sequence &
        Exfiltrate sensitive info & --- & Linux\\[1ex]
        \hline
        3CX & (1) Install trojanized software; (2) Run downloader; (3) Receive C2 servers; (4) Download third stage dataminer; (5) Steal browser info & setup.exe (upon installation) & Obfuscation, DLL side-loading, process injection & Steal data & ICONICSTEALER, DAVESHELL, SIGFLIP, VEILEDSIGNAL & WIN\\[1ex]
        \hline
        CloudEX &(1) Authenticate to exposed cloud service (initial access);
        (2) Discover residual CI/CD credentials;
        (3) Exploit credentials to access internal artifact repository;
        (4) Modify artifacts to embed downloader;
        (5) Trigger multi-stage malware via downstream build & CI/CD pipelines & Obfuscation & Steal data & Medusa, Nmap & WIN\\[1ex]
        \hline
        LayerInj & (1) Introduce backdoored model with conditional trigger;
        (2) Deploy in downstream service (Docker);
        (3) Inference triggers outbound C2 connection (class-match);
        (4) Exfiltrate data via activated payload & conditional trigger & Logic obfuscation, fileless malware & Steal data & Docker,Medusa & Linux\\
        \bottomrule
    \end{tabular}
\end{table*}

\paragraph{Scenario Design.}
The scenarios are constructed to reflect realistic attacker behavior across different ecosystems. Specifically, they include:
\textbf{package-based attacks} (e.g,\textit{Stegano}, \textit{Starter}, \textit{Parallel}, \textit{NPMEX}), where malicious logic is introduced via software dependencies and executed during installation or runtime; 
(ii) \textbf{software supply-chain attacks} (e.g., \textit{3CX}), involving trojanized binaries and multi-stage payload delivery; and 
(iii) \textbf{cloud and container-based attacks} (e.g., \textit{CloudEx}, \textit{LayerInj}), which exploits leaked credentials caused by misconfigurations and insufficient auditing of backdoored models.
These designs follow the exploitation trends identified through statistical analysis in Section~\ref{app:malicious package statistic analysis table}.

\paragraph{Execution Patterns.}
Across scenarios, we observe three common execution patterns:
(1) \textit{trigger-based execution}, where attacks are activated by specific events such as \texttt{setup.py} or installation hooks;
(2) \textit{multi-stage workflows}, where payloads are fetched and executed incrementally; and 
(3) \textit{parallel or decoupled execution}, where malicious components operate independently to evade detection.

\paragraph{Attack Flow Illustration}

\begin{figure}[t]
    \centering
    \footnotesize
    \includegraphics[width=1.0\linewidth]{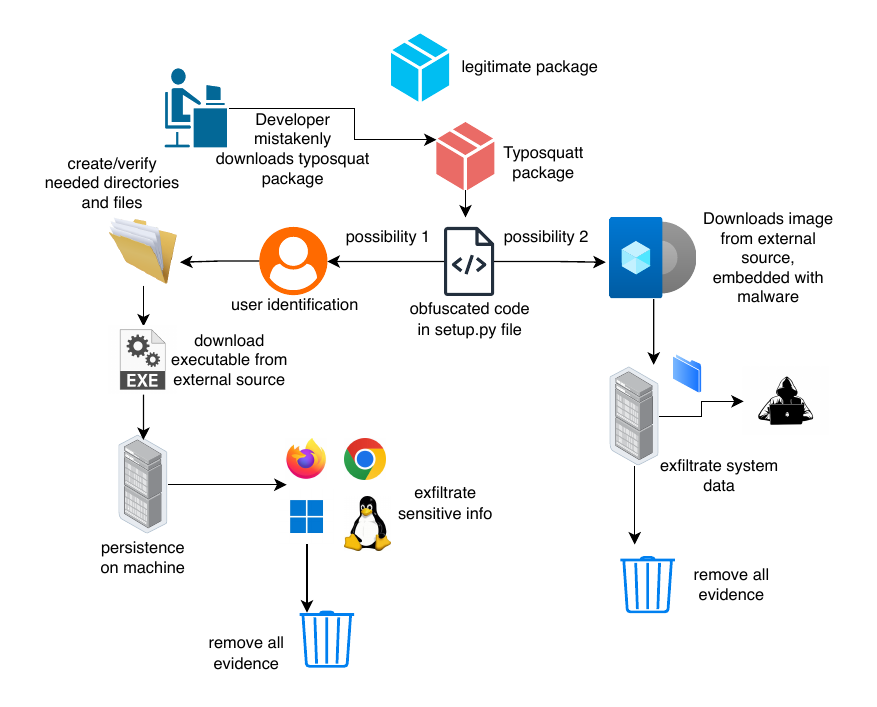}
    \caption{Stegano and Starter Attack Flow}
    \label{fig:attack scenario 1 and 2}
\end{figure}

\begin{figure}[t]
    \centering
    \footnotesize
    \includegraphics[width=\linewidth]{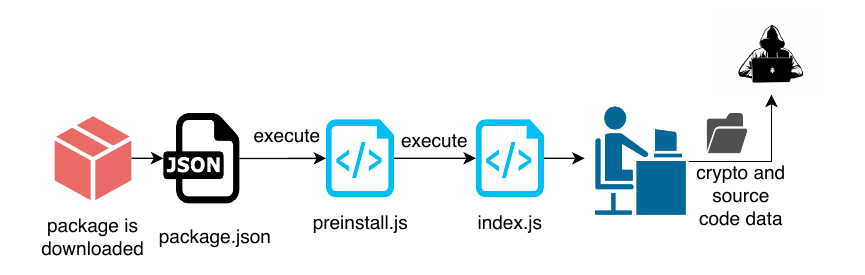}
    \caption{Parallel Attack Flow}
    \label{fig:attack scenario 3}
\end{figure}

\begin{figure}[t]
    \centering
    \footnotesize
    \includegraphics[width=0.8\linewidth]{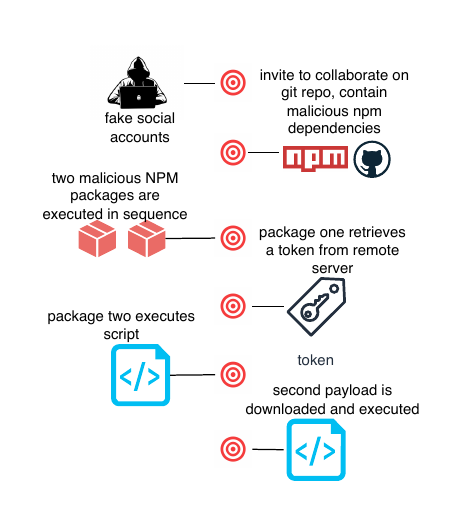}
    \caption{NPMEX Attack Flow}
    \label{fig:attack scenario 4}
\end{figure}

\begin{figure}[t]
    \centering
    \footnotesize
    \includegraphics[width=\linewidth]{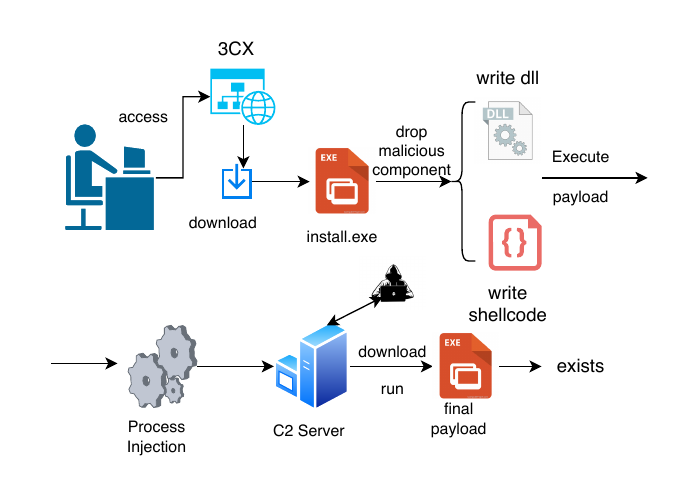}
    \caption{3CX Attack Flow}
    \label{fig:attack scenario 5}
\end{figure}

\begin{figure}[t]
    \centering
    \footnotesize
    \includegraphics[width=\linewidth]{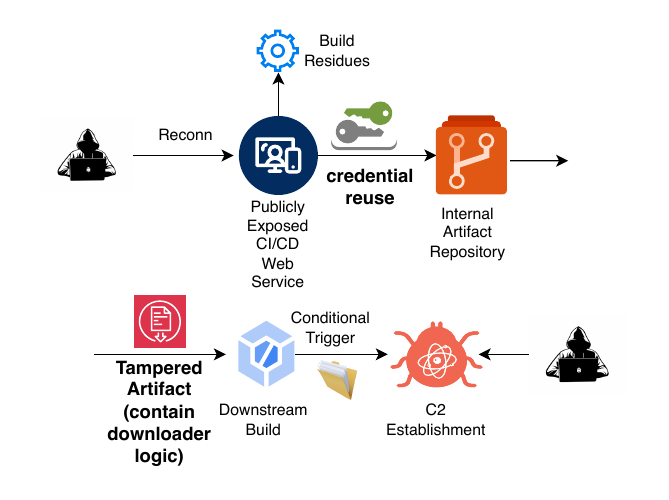}
    \caption{CloudEX Attack Flow}
    \label{fig:attack scenario 6}
\end{figure}

\begin{figure}[t]
    \centering
    \footnotesize
   \includegraphics[width=0.9\linewidth]{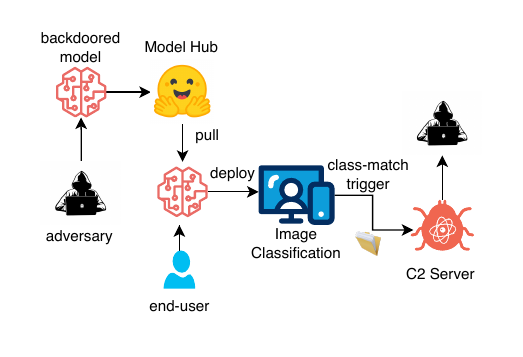}
    \caption{LayerInj Attack Flow}
    \label{fig:attack scenario 7}
\end{figure}

To provide concrete insights into these behaviors, Figures~\ref{fig:attack scenario 1 and 2}--\ref{fig:attack scenario 7} visualize representative attack flows. 
Each diagram highlights the sequence of actions, the interaction between components, and the points where malicious behaviors are triggered. 
For example, Figure~\ref{fig:attack scenario 3} demonstrates a parallel execution model in which multiple scripts are triggered via installation hooks, while Figure~\ref{fig:attack scenario 5} shows a multi-stage supply-chain attack involving downloader and data exfiltration components.

\subsection{IOC Coverage and Data Quality Discussion}
\label{app:ioc coverage and data quality}

The 2{,}919 verified IOC records (Table~\ref{tab:ioc-gt}) do not claim exhaustive coverage of every attack action. Across the seven scenarios, we document 47 discrete attack steps (Table~\ref{tab:ioc-coverage})from the operator's perspective (based on the recorded attack timelines) and verify that \textbf{37 (78.7\%)} produce at least one IOC record in the collected telemetry. The remaining 10 steps --- all attributable to \emph{in-memory execution} via the C2
framework (e.g., Mythic's \texttt{load\_script}, or interactive shell commands) or encrypted payload content --- leave no host-level or file-system artifact by design. 

These results indicate phase-level IOC presence, not end-to-end causal
chain reconstructability. In other words, a phase may leave at least one
validated IOC while still lacking the provenance links required to connect
it to adjacent phases.
The deficit concentrates in \emph{post-exploitation} phase (Discovery 33\%, Collection 44\%), where the attacker deliberately operates in memory to avoid disk and process-creation artifacts. This gap is not a data-collection failure but a faithful reproduction of the observability boundary that real-world defenders face: network-layer sensors (Zeek)
capture every TCP/UDP flow including timing, volume, and port usage,
while host sensors (Sysmon, syslog) record process creation and file
events, but neither can inspect the content of in-memory-only C2
commands. Even for unobservable steps, \emph{indirect evidence}
persists---for instance, data-volume spikes in \texttt{zeek\_conn}
coincide with documented exfiltration times, and an anomalous 15{,}068-byte
HTTP response in sc7 corresponds to the \texttt{load\_script} upload---
providing analysts with inferential signals despite the absence of
explicit IOC records.

\section{LLM-Assisted TTPs Analysis}
\label{app:llm-assisted analysis}

This appendix details the LLM-assisted pipeline used to extract
ATT\&CK technique mappings from payload source code, including
the prompt design, automated post-processing, and human validation
protocol summarized in Section~\ref{sec:methodology}.

\paragraph{LLM-Assisted TTPs Extraction}

For payload-originated behavious --- where no structured C2 tasking log is available --- we employ GPT-5.1 to generate candidate ATT\&CK
technique mappings directly from payload source code. The model is
prompted with a role-based instruction to act as a cybersecurity
analyst and identify all techniques that are \emph{referenced,
implied, or likely used} in the provided code. This deliberately broad scope prioritizes \textbf{recall} at the candidate-generate stage; precision is ensured through the human validation described next. The model returns a structured JSON array, with each entry containing
a technique ID and its associated tactic. We then apply automated
post-processing to extract the JSON output and discard malformed
entries missing required fields.

\paragraph{Human Validation.}
Each candidate technique produced by the pipeline underwent a
two-phase review conducted independently by two researchers with
adversary-emulation experience. In the \emph{precision review},
annotators examined the payload source code to verify whether each
candidate technique corresponded to an identifiable behavior in
the code, such as specific API calls (e.g., system command invocations, or recognizable code patterns; candidates without traceable evidence in the payload were removed as false positives. In the \emph{recall review},
annotators cross-referenced the payload code against the recorded
attack steps of each scenario to identify techniques that the LLM
failed to surface. Disagreements between the two annotators were
resolved through discussion until consensus was reached; most
divergences involved ambiguous sub-technique granularity or
dual-purpose code patterns.

\paragraph{LLM Prompt Template}

The following prompt template is used to query GPT-5.1 for candidate
ATT\&CK technique extraction. The placeholder \texttt{\{SOURCE\_CODE\}}
is replaced with the full payload source code at inference time.

\begin{lstlisting}[
  basicstyle=\footnotesize\ttfamily,
  frame=single,
  breaklines=true,
  caption={Prompt template for TTP extraction.},
  label={lst:prompt}
]
You are a cybersecurity analyst. Given the following source code or logs, extract all MITRE ATT&CK techniques referenced, implied, or likely used.

Return JSON with fields:
[
  {"techniqueID": "Txxxx" or "Txxxx.yyy",
   "tactic": "<tactic-name>"},
  ...
]

Source:
{SOURCE_CODE}
\end{lstlisting}

The model returns a JSON array where each element contains a \texttt{techniqueID} (e.g., \texttt{T1059.001}) and a \texttt{tactic} field (e.g., \texttt{execution}); malformed entries missing either field are discarded in post-processing.

Upon receiving the LLM response, we apply the following automated
post-processing steps:
\begin{enumerate}
    \item \textbf{JSON extraction.} A regular expression is used to
        locate the first JSON array in the raw LLM output, handling
        cases where the model produces preamble text or markdown
        formatting around the JSON.
    \item \textbf{Field validation.} Each parsed element is checked
        for the presence of both \texttt{techniqueID} and
        \texttt{tactic} fields; entries missing either field are
        discarded.
    \item \textbf{Human review.} The filtered candidates are passed
        to two independent annotators for precision and recall
        validation.
\end{enumerate}

\section{Cross-Scenario ATT\&CK Technique Distribution}
\label{app:ttp-distribution}

We analyze how ATT\&CK techniques are distributed across scenarios to characterize both shared behavioral patterns and scenario-specific specializations. 
Specifically, we consider (i) the breadth of techniques within each scenario and (ii) the prevalence of techniques across scenarios.

\paragraph{Technique Breadth and Uniqueness}
Table~\ref{tab:scenario_ttp_stats} summarizes scenario-level statistics. 
We observe substantial variation in technique breadth, where $|\mathcal{T}_s|$ counts distinct techniques observed in scenario $s$ and $|\mathcal{T}|$ is the union over all scenarios; $f$ denotes cross-scenario frequency.
SC1/SC2/SC6/SC7 cover roughly two-thirds of the global technique pool, whereas SC3--SC5 are much narrower in total techniques but exhibit a markedly higher fraction of scenario-unique techniques.
This pattern suggests that broad scenarios share a large common core of behaviors, while narrower scenarios emphasize more specialized, scenario-specific steps.
Complementarily, we rank techniques by cross-scenario coverage and find a small set of ubiquitous techniques that appear in almost all scenarios (Table~\ref{tab:top_techniques}).

Notably, supply-chain--related techniques are consistently present in our scenario set, reflecting that several scenarios are dependency- or software-distribution--mediated intrusions. The most prevalent techniques largely correspond to capability acquisition/staging and tool transfer, consistent with supply-chain or dependency-mediated intrusion setups where attackers must first obtain and deliver artifacts before executing later-stage actions.

\begin{table*}[t]
\centering
\footnotesize
\caption{Scenario-level ATT\&CK technique coverage and uniqueness}
\label{tab:scenario_ttp_stats}
\begin{tabular}{@{}lccccccc@{}}
\toprule
 & \textbf{SC1} & \textbf{SC2} & \textbf{SC3} & \textbf{SC4} & \textbf{SC5} & \textbf{SC6} & \textbf{SC7} \\
\midrule
Total observed techniques $|\mathcal{T}_s|$ & 104 & 103 & 29 & 29 & 35 & 104 & 100 \\
Coverage of global pool ($|\mathcal{T}|=161$) & 64.6\% & 64.0\% & 18.0\% & 18.0\% & 21.7\% & 64.6\% & 62.1\% \\
Scenario-unique techniques ($f=1$) & 7 & 6 & 9 & 7 & 13 & 6 & 5 \\
Unique share within scenario & 6.7\% & 5.8\% & 31.0\% & 24.1\% & 37.1\% & 5.8\% & 5.0\% \\
\bottomrule
\end{tabular}
\end{table*}

\paragraph{Cross-Scenario Prevalence.}

Complementing the above, Table~\ref{tab:top_techniques} lists the most common techniques ranked by the number of scenarios in which they appear.
We observe a small set of highly prevalent techniques that occur in nearly all scenarios, forming a \emph{shared behavioral backbone} across attacks.

\begin{table}[t]
\centering
\footnotesize
\caption{Top-10 most common ATT\&CK techniques ranked by scenario coverage.
Six techniques appear in all 7 scenarios, two appear in 6 scenarios; the remaining techniques are tied at 5 scenarios, from which we report two representative examples.}
\label{tab:top_techniques}
\begin{tabular}{@{}llr@{}}
\toprule
\textbf{Technique ID} & \textbf{Name (MITRE ATT\&CK)} & \textbf{\#Scenarios} \\
\midrule
T1082 & System Information Discovery & 7 \\
T1083 & File and Directory Discovery & 7 \\
T1105 & Ingress Tool Transfer & 7 \\
T1195.002 & Compromise Software Supply Chain & 7 \\
T1588 & Obtain Capabilities & 7 \\
T1608 & Stage Capabilities & 7 \\
\midrule
T1033 & System Owner/User Discovery & 6 \\
T1070.004 & Indicator Removal: File Deletion & 6 \\
T1005 & Data from Local System & 5 \\
T1012 & Query Registry & 5 \\
\bottomrule
\end{tabular}

\caption*{\footnotesize\textbf{Note:} In addition to the techniques listed above, 18 techniques are tied with coverage in 5 scenarios (see Supplementary).}
\end{table}

Notably, several of these techniques are related to capability acquisition, staging, and tool transfer (e.g., ingress of payloads and preparation of execution artifacts), which are fundamental steps in multi-stage attacks. 
Supply-chain--related techniques are also consistently present, reflecting that multiple scenarios involve dependency or software distribution as an initial access vector.

\paragraph{Summary.}
Together, these results indicate a two-level structure in the attack space: a stable core of widely shared techniques that underpin most scenarios, and a set of scenario-specific techniques that capture unique behaviors. 
This structure motivates evaluating detection approaches under both common and specialized technique distributions.

\section{Full Scenario Result}
\label{app:scenario result}

\begin{table*}[t]
\centering
\small
\caption{Per-scenario best-achievable reconstruction under each source budget (selected by maximizing StepR; tie-break by ChainR then event volume). Each cell reports the selected source\_set, StepR, and missing expected steps. Abbrev: E\_s=Expected\_Steps, I=INSTALL, D=DOWNLOAD, O=OUTBOUND\_CONN, E=EXFIL, A=AUTH.}
\label{tab:scenario_budget_breakdown}
\setlength{\tabcolsep}{4pt}
\renewcommand{\arraystretch}{1.08}

\newcommand{\src}[1]{\texttt{\small #1}}

\begin{tabular}{@{}l c p{3.4cm} p{3.4cm} p{6.0cm}@{}}
\toprule
Scenario & $E_s$ & Single (best) & Combo (best) & Multi (full telemetry) \\
\midrule

SC1 & 4 &
\makecell[l]{\src{azure\_process}\\[-1pt]\scriptsize StepR=0.250; miss=\{D,E,O\}} &
-- &
\makecell[l]{\src{azure\_conn+\allowbreak azure\_process+}\\
\src{azure\_security+\allowbreak azure\_events+\allowbreak azure\_port}\\\scriptsize StepR=0.250; miss=\{D,E,O\}} \\

SC2 & 4 &
\makecell[l]{\src{syslog}\\[-1pt]\scriptsize StepR=0.500; miss=\{D,O\}} &
-- &
\makecell[l]{\src{azure\_events+\allowbreak syslog}\\[-1pt]\scriptsize StepR=0.750; miss=\{O\}} \\

SC3 & 4 &
\makecell[l]{\src{suricata}\\[-1pt]\scriptsize StepR=0.250; miss=\{D,E,I\}} &
\makecell[l]{\src{zeek+\allowbreak syslog}\\[-1pt]\scriptsize StepR=0.500; miss=\{D,E\}} &
\makecell[l]{\src{auditd+\allowbreak auth+\allowbreak suricata+\allowbreak syslog+\allowbreak zeek}\\[-1pt]\scriptsize StepR=0.500; miss=\{D,E\}} \\

SC4 & 4 &
\makecell[l]{\src{syslog}\\[-1pt]\scriptsize StepR=0.500; miss=\{E,O\}} &
\makecell[l]{\src{zeek+\allowbreak syslog}\\[-1pt]\scriptsize StepR=0.750; miss=\{E\}} &
\makecell[l]{\src{auditd+\allowbreak auth+\allowbreak suricata+\allowbreak syslog+\allowbreak zeek}\\[-1pt]\scriptsize StepR=0.750; miss=\{E\}} \\

SC5 & 4 &
-- & -- &
\makecell[l]{\src{azure\_events}\\[-1pt]\scriptsize StepR=0.250; miss=\{D,E,O\}} \\

SC6 & 4 &
\makecell[l]{\src{syslog}\\[-1pt]\scriptsize StepR=0.250; miss=\{D,E,O\}} &
-- &
\makecell[l]{\src{azure\_events+\allowbreak syslog}\\[-1pt]\scriptsize StepR=0.250; miss=\{D,E,O\}} \\

SC7 & 3 &
\makecell[l]{\src{suricata}\\[-1pt]\scriptsize StepR=0.667; miss=\{D\}} &
\makecell[l]{\src{auditd+\allowbreak zeek}\\[-1pt]\scriptsize StepR=0.667; miss=\{D\}} &
\makecell[l]{\src{auditd+\allowbreak suricata+\allowbreak syslog+\allowbreak zeek+\allowbreak tracee}\\[-1pt]\scriptsize StepR=0.667; miss=\{D\}} \\

\bottomrule
\end{tabular}
\end{table*}

We note that the \textbf{Multi (full telemetry)} column enumerates the complete telemetry inventory per scenario.
We treat \texttt{events} as a composite multi-channel source (aggregated at export) and thus do not count \texttt{events}-based settings as single-source in budget accounting.

\paragraph{Interpretation of missing anchors.}
The missing-step entries in Table~14 denote anchors not recovered by our conservative rule-based reconstruction, not necessarily the absence of any supporting raw evidence. We therefore use these entries as the audit set for the validation analysis in Appendix~\ref{app:csa3_details}.

\section{Cross-Scenario Analysis Details}
\label{app:cross_scenario_details}

This appendix expands the mechanism-based cross-scenario analysis summarized in Section \ref{sec:cross-scenario analysis}. We provide (i) a reconstructability typing that explains dominant success/failure patterns, (ii) alignment and evidence-conditioned detectability context, (iii) a failure taxonomy with diagnostic value, and (iv) deployment implications derived from the taxonomy.

\subsection{CSA-1 Details: Mechanism-Based Typing for Reconstructability}
\label{app:csa1_details}

Across scenarios, end-to-end reconstruction depends on whether telemetry provides: (i) \emph{phase anchors}---events that unambiguously indicate a coarse step such as \texttt{INSTALL}, \texttt{DOWNLOAD}, \texttt{\seqsplit{OUTBOUND\_CONN}}, or \texttt{EXFIL}; and (ii) \emph{joinable identifiers}---stable entities that allow anchors to be chained (host/user/process identifiers, network endpoints, workload IDs, and consistent timestamps).

\paragraph{Type I: Joinable host--network chains (reconstructable).}
In this type, host activity and network activity are both visible and can be linked. Reconstruction succeeds because the pipeline can (a) anchor installation and code execution on provenance/audit or process traces, and (b) attribute outbound connections to the same process/host context. The resulting narrative is robust even under conservative correlation parameters, since multiple anchors corroborate one another (e.g., a download event followed by process execution and a temporally nearby outbound session).

\paragraph{Type II: Evidence-present but weakly joinable (partially reconstructable).}
Here, phases may be present but linkage is fragile. Common causes include missing process-to-socket attribution (network flows exist but cannot be tied to a process), non-unique entities (shared hosts/users across multiple parallel tasks), and temporal collision under concurrent benign activity. The pipeline may observe the correct step set but cannot confidently choose a single causal path, so it favors shorter chains over brittle long chains. This behavior is desirable from a threat-hunting perspective: it avoids over-claiming end-to-end completion when the evidence does not support a unique narrative.

\paragraph{Type III: Structural observability gaps (bounded reconstructability).}
In this type, at least one expected phase is absent under the available schemas and collection boundary. Examples include control-plane actions occurring outside the host boundary, event exports lacking fields needed to distinguish download vs. generic file activity, or missing attribution keys that prevent linking host and cloud actions. Multi-source correlation cannot recover phases that are never observed; improvements require targeted telemetry that directly exposes the missing phase and provides join keys (e.g., cloud IAM/API audit logs with request IDs, egress proxy logs, or high-fidelity tracing for process/network attribution).

\paragraph{Implications.}
This typing clarifies why adding more sources does not automatically yield full chains: \emph{evidence diversity only helps if it adds missing anchors or strengthens joins}. In practice, the most impactful additions are those that expose currently missing phases (Type III) or provide stable join keys for phases that are otherwise isolated (Type II).

\subsection{CSA-2 Details: Alignment, Observability, and Evidence-Conditioned Feasibility}
\label{app:csa2_details}

We align reconstructed coarse steps to MITRE ATT\&CK techniques post hoc. This section clarifies why alignment quality varies across scenarios even under identical pipeline parameters.

\begin{figure*}[t]
\centering
\caption{Scenario-level comparison of ATT\&CK coverage and time window: \\ 
number of techniques, number of tactics, and attack duration.}
\small
\begin{minipage}{0.32\textwidth}
\centering
\begin{tikzpicture}
\begin{axis}[
  xbar, xmin=0,
  width=\linewidth, height=5.2cm,
  xlabel={\#Techniques},
  symbolic y coords={Starter,Stegano,CloudEX,LayerInj,3CX,NPMEX,Parallel},
  ytick=data, y dir=reverse,
  bar width=3.5pt,
  axis lines=left,
  enlarge y limits=0.08
]
\addplot coordinates {
  (103,Starter) (104,Stegano) (104,CloudEX) (100,LayerInj) (35,3CX) (29,NPMEX) (29,Parallel)
};
\end{axis}
\end{tikzpicture}
\end{minipage}\hfill
\begin{minipage}{0.32\textwidth}
\centering
\begin{tikzpicture}
\begin{axis}[
  xbar, xmin=0, xmax=14,
  width=\linewidth, height=5.2cm,
  xlabel={\#Tactics},
  symbolic y coords={Starter,Stegano,CloudEX,LayerInj,3CX,NPMEX,Parallel},
  ytick=data, y dir=reverse,
  bar width=3.5pt,
  axis lines=left,
  enlarge y limits=0.08
]
\addplot coordinates {
  (14,Starter) (14,Stegano) (14,CloudEX) (14,LayerInj) (8,3CX) (11,NPMEX) (8,Parallel)
};
\end{axis}
\end{tikzpicture}
\end{minipage}\hfill
\begin{minipage}{0.32\textwidth}
\centering
\begin{tikzpicture}
\begin{axis}[
  xbar, xmin=0,
  width=\linewidth, height=5.2cm,
  xlabel={Duration (min)},
  symbolic y coords={Starter,Stegano,CloudEX,LayerInj,3CX,NPMEX,Parallel},
  ytick=data, y dir=reverse,
  bar width=3.5pt,
  axis lines=left,
  enlarge y limits=0.08
]
\addplot coordinates {
  (330,Starter) (121,Stegano) (71,CloudEX) (37,LayerInj) (18,3CX) (5,NPMEX) (17,Parallel)
};
\end{axis}
\end{tikzpicture}
\end{minipage}

\label{fig:scenario_bars}
\end{figure*}

\paragraph{Heterogeneous projections of the same action.}
The same underlying attacker action may manifest as different observables depending on the source: a retrieval phase can appear as a package-manager transaction, a file creation event, or a network flow. Alignment therefore becomes evidence-conditioned: if the dataset lacks the projection that carries sufficient context (e.g., process attribution, URL/domain, or artifact lineage), the corresponding technique may be under-supported despite being executed.

\paragraph{Abstraction gap: step-level evidence vs. technique-level diversity.}
Coarse steps intentionally compress technique diversity to preserve cross-scenario comparability. This trades granularity for interpretability: the pipeline is optimized for reconstructing \emph{chain structure} rather than identifying exact technique variants. As a result, alignment is best read as an ``evidence supports this phase'' signal, not as a direct technique detector. This is especially relevant for scenarios whose maliciousness is semantic rather than system-level (e.g., model-level backdoors), where OS telemetry may show standard inference workflows while the malicious effect appears only in model outputs.

\paragraph{Scenario-level breadth and temporal window.}
Scenario breadth (number of techniques/tactics) and attack window duration modulate alignment difficulty. Long windows increase background activity and temporal collisions; short windows can compress transitions and obscure intermediate anchors. Figure~\ref{fig:scenario_bars} visualizes this diversity and motivates why a single alignment strategy must remain conservative.

\subsubsection{Evidence-Conditioned Detectability Matrix (Context, Not Accuracy)}
\label{app:detectability_matrix_details}

Table~\ref{tab:scenario_method_matrix_binary_no_baseline} provides a binary feasibility view: whether the evidence a detector family requires is typically observable for each scenario. This matrix is complementary to Table~\ref{tab:source-budget-summary}: it is not a measured accuracy claim, but a structured explanation of why certain detector families are ill-suited under evidence constraints.

Two cross-cutting observations follow. First, methods that rely on stable indicators (IOC/signatures) are systematically brittle across supply-chain scenarios because artifacts are often novel, polymorphic, or ephemeral, and the dominant signal is behavioral rather than static. Second, methods that require a particular evidence boundary (e.g., host-only provenance or cloud-only integrity checks) fail when the core exploit occurs outside that boundary. Our multi-source chaining approach is feasible across all scenarios because it composes whichever evidence is present into a unified narrative; however, feasibility does not imply completeness, and bounded observability (CSA-1 Type III) remains a limiting factor.

\begin{table*}[t]
\centering
\caption{Evidence-conditioned detectability matrix (binary feasibility)}
\footnotesize
\setlength{\tabcolsep}{2.9pt}
\renewcommand{\arraystretch}{1.18}
\begin{tabular*}{\textwidth}{@{\extracolsep{\fill}}%
C{3.3cm}      
C{1.5cm}
C{1.8cm}
C{1.2cm}
C{1.4cm}
C{1.2cm}
C{1.25cm}
C{1.25cm}
C{1.45cm}
C{1.40cm}
@{}}
\hline
\textbf{Scenario (core exploit)} &
\makecell{\textbf{IOC}\\\textbf{/ Sig.}} &
\makecell{\textbf{SCA}\\\textbf{+SBOM}} &
\makecell{\textbf{Behavior}\\\textbf{+ATT\&CK}} &
\makecell{\textbf{1-class}\\\textbf{Anom.}} &
\makecell{\textbf{Call/}\\\textbf{Syscall}\\\textbf{Graphs}} &
\makecell{\textbf{Single-src}\\\textbf{Prov.}} &
\makecell{\textbf{Cross-src}\\\textbf{Corr.}} &
\makecell{\textbf{Model}\\\textbf{Integrity}} &
\textbf{Ours} \\
\hline

\textbf{(Sources; core idea $\rightarrow$)} &
\makecell{\footnotesize hashes/YARA/\\domains;\footnotesize match \\known-bad \\ \cite{10.1016/j.jisa.2024.103738}} &
\makecell{\footnotesize repo/build/\\\footnotesize SBOM;\\\footnotesize dep/prov \\anomalies \\ \cite{10.1145/3766073,10352114}} &
\makecell{\footnotesize EDR/auditd;\\\footnotesize stage/seq. \\rules \\ \cite{SUN2024103772}} &
\makecell{\footnotesize features;\\\footnotesize outlier \\scoring \\ \cite{Omar2022}} &
\makecell{\footnotesize eBPF/audit/\\traces;\\\footnotesize sequence \\graphs \\\cite{10254962}} &
\makecell{\footnotesize OS prov.;\\\footnotesize causal \\ \footnotesize chain \\\cite{10.1145/3539605}} &
\makecell{\footnotesize host+net;\\\footnotesize join \\evidence \\\cite{LOTLHunter2026,APTMCL2026}} &
\makecell{\footnotesize model registry\\+eval;\\\footnotesize attest+\\trigger tests \\\cite{10.5555/3766078.3766404}} &
\makecell{\footnotesize host+net+\\\footnotesize proc.+tracee;\\\footnotesize unified corr.\\ + \footnotesize causal \\\footnotesize chaining} \\
\hline

SC1-Stegano (steganography) &
-- & -- & \checkmark & \checkmark & \checkmark & \checkmark & \checkmark & -- & \checkmark \\
SC2-Starter (autostart) &
-- & -- & \checkmark & \checkmark & \checkmark & \checkmark & \checkmark & -- & \checkmark \\
SC3-Parallel (multi-stage) &
-- & \checkmark & \checkmark & -- & \checkmark & \checkmark & \checkmark & -- & \checkmark \\
SC4-NPMEX (dependency chain) &
-- & \checkmark & \checkmark & -- & \checkmark & \checkmark & \checkmark & -- & \checkmark \\
SC5-3CX (plugin, multi-stage backdoor) &
-- & -- & \checkmark & -- & \checkmark & \checkmark & \checkmark & -- & \checkmark \\
SC6-CloudEX (leaked cloud credential) &
-- & \checkmark & \checkmark & -- & -- & -- & -- & -- & \checkmark \\
SC7-LayerInj (backdoored model) &
-- & -- & -- & -- & -- & -- & -- & \checkmark & \checkmark \\
\hline
\textbf{Per-method proportion ($\checkmark/7$)} &
\textbf{0/7} &
\textbf{3/7} &
\textbf{6/7} &
\textbf{2/7} &
\textbf{5/7} &
\textbf{5/7} &
\textbf{5/7} &
\textbf{1/7} &
\textbf{7/7} \\
\hline
\end{tabular*}
\label{tab:scenario_method_matrix_binary_no_baseline}
\end{table*}

\subsection{CSA-3 Details: Failure Taxonomy and Diagnostic Value}
\label{app:csa3_details}

Table~\ref{tab:scenario_method_matrix_binary_no_baseline} already characterizes \emph{feasibility} under evidence availability. Here we add only a lightweight \emph{diagnostic} view, summarized in Table~\ref{tab:failure_minimal}, that is specific to our pipeline outputs: when reconstruction fails, it typically manifests as (i) \textbf{missing-phase gaps} (expected steps never observed under the collected schemas), (ii) \textbf{attribution breaks} (steps observed but not joinable across sources into a unique chain), or (iii) \textbf{negative/partial chains} (attempt signals without downstream execution/connectivity). We report these gaps explicitly via missing-step diagnostics (e.g., \texttt{MISSING\_*}, \texttt{NO\_STEPS\_OBSERVED}, \texttt{PREFILTER\_UNUSABLE}) to distinguish evidence absence from schema/rule mismatch and to avoid over-claiming end-to-end compromise.

\begin{table}[t]
\centering
\footnotesize
\caption{Minimal diagnostic view of reconstruction failures.}
\label{tab:failure_minimal}
\begin{tabular}{@{}l@{\hspace{4pt}}l@{\hspace{4pt}}l@{}}
\toprule
\textbf{Category} & \textbf{Symptom} & \textbf{Typical remedy} \\
\midrule
Missing-phase gap & \texttt{MISSING\_*} persists & add phase-specific telemetry \\
Attribution break & steps present, chain weak & add/join stable identifiers \\
Negative/partial chain & attempt w/o completion & require downstream anchors \\
\bottomrule
\end{tabular}
\end{table}

\paragraph{Validation checks.}
We add two validation checks to clarify the interpretation of missing anchors under our conservative rule-based reconstruction.
First, for each missing \texttt{DOWNLOAD}, \texttt{EXFIL}, and \texttt{OUTBOUND\_CONN} anchor in Table~\ref{tab:scenario_budget_breakdown}, we manually inspect only defender-visible victim-side raw telemetry, excluding attacker-side logs and C2/operator records.
Each case is classified as evidence absent, present but non-joinable, conservative-rule miss, or semantic ambiguity.
Evidence absent means no supporting victim-side raw evidence is found; present but non-joinable means evidence exists but lacks stable process, user, session, flow, endpoint, or file-lineage identifiers; conservative-rule miss means raw evidence exists but is not captured by the portable rule tagger; semantic ambiguity means the evidence is suggestive but insufficient for stable phase assignment without semantic interpretation.
Second, we compare first-step-only candidate windows against downstream-context candidate windows as a candidate-volume proxy.

\begin{table*}[!htbp]
\centering
\caption{Manual audit of missing anchors. The audit uses only defender-visible victim-side telemetry.}
\label{tab:missing_anchor_audit}
\scriptsize
\begin{tabularx}{\textwidth}{llcclX}
\toprule
Scenario & Missing anchor & Raw evidence & Joinable & Final category & Evidence basis \\
\midrule
SC1 & \texttt{DOWNLOAD} & Yes & Yes & Conservative-rule miss & Azure connection records contain Python network traffic matching the attack infrastructure, but the portable tagger does not recover it as a \texttt{DOWNLOAD} anchor. \\
SC1 & \texttt{OUTBOUND\_CONN} & Yes & Yes & Conservative-rule miss & Azure connection records expose outbound Python traffic to the attack infrastructure, but it is not recovered by the conservative step rules. \\
SC1 & \texttt{EXFIL} & Yes & Yes & Semantic ambiguity & Network evidence is present, but direction/content semantics are insufficient to stably assign an \texttt{EXFIL} anchor under portable rules. \\
SC2 & \texttt{OUTBOUND\_CONN} & Yes & Yes & Conservative-rule miss & Azure event records contain joinable runtime/external-connection evidence, but the conservative tagger misses the outbound anchor. \\
SC3 & \texttt{DOWNLOAD} & Yes & Yes & Conservative-rule miss & Suricata/Eve records contain attack-related retrieval evidence, but it is not labeled as a portable \texttt{DOWNLOAD} anchor. \\
SC3 & \texttt{EXFIL} & Yes & Yes & Conservative-rule miss & Suricata/Eve records contain supporting transfer evidence, but the rule layer does not recover it as an \texttt{EXFIL} anchor. \\
SC4 & \texttt{EXFIL} & Yes & Yes & Semantic ambiguity & Suricata/Eve evidence is suggestive of transfer activity, but the available telemetry does not provide stable semantics for an \texttt{EXFIL} anchor. \\
SC5 & \texttt{DOWNLOAD} & Yes & Yes & Conservative-rule miss & Download URL, Zone.Identifier, and installer artifacts are present but not recovered by the conservative portable anchor rules. \\
SC5 & \texttt{OUTBOUND\_CONN} & Yes & Yes & Conservative-rule miss & A joinable external connection is observed from the X\_TRADER installation chain: \texttt{msiexec.exe /i X\_TRADER.msi} connects to \texttt{192.124.249.36}/\texttt{cloudproxy10036.sucuri.net:80}; C2 semantics are not claimed. \\
SC5 & \texttt{EXFIL} & Partial & No & Semantic ambiguity & External connections exist, but Sysmon network telemetry lacks byte-level, payload, upload, or file-transfer evidence sufficient to infer exfiltration. \\
SC6 & \texttt{DOWNLOAD} & Yes & Yes & Conservative-rule miss & Victim-side Azure events contain supporting download-related evidence, but the conservative tagger does not recover it as a portable anchor. \\
SC6 & \texttt{OUTBOUND\_CONN} & Yes & Yes & Conservative-rule miss & Victim-side Azure events contain outbound-connection evidence, but it is missed by the conservative rule layer. \\
SC6 & \texttt{EXFIL} & Yes & Yes & Semantic ambiguity & Supporting evidence exists, but available fields do not stably distinguish exfiltration from semantically related transfer activity. \\
SC7 & \texttt{DOWNLOAD} & Yes & Yes & Conservative-rule miss & Zeek file-transfer evidence is present and joinable, but the rule tagger misses it as a portable \texttt{DOWNLOAD} anchor. \\
\bottomrule
\end{tabularx}
\end{table*}

Overall, the audit classifies 10 of 14 missing anchors as conservative-rule misses and 4 as semantic ambiguities, with no case requiring attacker-side evidence. Thus, missing reconstruction anchors should be interpreted as lower-bound rule-matchable anchors, not exhaustive raw-log absence.

\begin{table}[!htbp]
\centering
\caption{First-step-only vs. downstream-context candidate analysis. Candidate windows are a benign-volume proxy, not a detector benchmark.}
\label{tab:candidate_reduction}
\scriptsize
\setlength{\tabcolsep}{3pt}
\begin{tabular}{@{}lrrrrr@{}}
\toprule
Setting & Benign ev. & Attack ev. & Benign win. & Attack win. & Retained SC \\
\midrule
First-step & 1541 & 1622 & 74 & 71 & 7/7 \\
Chain-context & 1012 & 1236 & 21 & 36 & 7/7 \\
\bottomrule
\end{tabular}
\end{table}

Adding downstream chain context reduces benign candidate windows from 74 to 21, a 71.6\% reduction, while retaining all seven attack scenarios.

\subsection{CSA-4: Sensitivity to Telemetry Density}
\label{app:csa4_full_sensitivity}

This appendix provides the complete results for the telemetry density sensitivity analysis summarized in Section~\ref{sec:csa4-sensitivity}.

\paragraph{Experiment design.}
For each telemetry source in a given configuration, we apply independent Bernoulli downsampling at rate $r \in \{1.0, 0.75, 0.5, 0.25, 0.1\}$ on raw parsed events \emph{before} step tagging.
For each rate, we execute 5 independent runs with seeds $\{0,1,2,3,4\}$ and aggregate using the same cross-scenario weighting scheme as Table~\ref{tab:source-budget-summary}: coverage and recall are weighted by expected steps ($E_s$), while precision and reconstructability use unweighted scenario means.

\begin{figure}[h]
  \centering
  \includegraphics[width=\columnwidth]{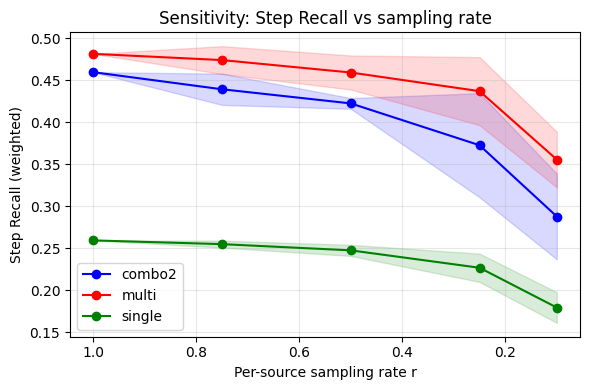}
  \caption{Step Recall (weighted) vs.\ per-source sampling rate by source budget.}
  \label{fig:sens-stepR}
\end{figure}
 
\begin{figure}[h]
  \centering
  \includegraphics[width=\columnwidth]{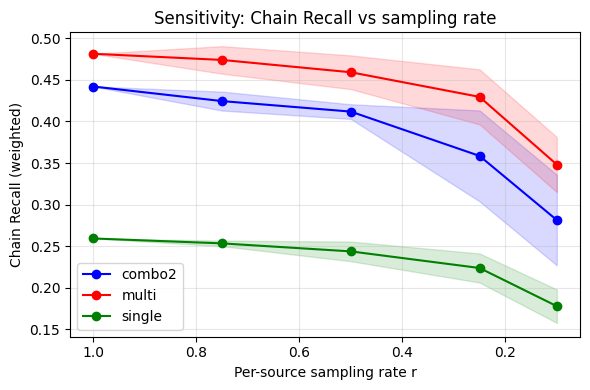}
  \caption{Chain Recall (weighted) vs.\ per-source sampling rate by source budget.}
  \label{fig:sens-chainR}
\end{figure}

\paragraph{Per-metric degradation curves.}
Figures~\ref{fig:sens-stepR} and~\ref{fig:sens-chainR} show Step Recall (weighted) and Chain Recall (weighted) as a function of sampling rate, complementing the Reconstructability plot in Figure~\ref{fig:sens-recon}.

Across all three metrics, the ordering single $<$ combo $<$ multi is preserved at every sampling rate, and multi-source configurations consistently exhibit the smallest relative performance drop.
Step Recall and Chain Recall track each other closely, as expected from the high step-to-chain precision observed in the full-fidelity experiments (Table~\ref{tab:source-budget-summary}).
 
\paragraph{Discussion.}
The results confirm that the multi-source advantage
reported in Table~\ref{tab:source-budget-summary} is robust under further evidence reduction: even under $4\times$ downsampling ($r{=}0.25$), multi-source configurations retain a higher fraction of their baseline performance than single-source settings.
This robustness arises from the complementary-anchor mechanism identified in CSA-1: when sampling removes events from one stream, alternative anchors in other streams preserve phase coverage and joinability.
However, at extreme sparsity ($r{=}0.1$), all configurations converge toward lower performance, indicating a floor below which no source-budget strategy can compensate for pervasive evidence loss.

\subsection{CSA-5: Structural Patterns \& Deployment Implications}
\label{app:csa5_deployment_implications}

\begin{table*}[t]
\centering
\caption{Reconstruction diagnostics for SC4 and SC1 derived from the evidence packages.}
\small
\begin{tabular}{l p{0.2\linewidth} p{0.15\linewidth} p{0.12\linewidth} r r r}
\hline
\textbf{Scenario} & \textbf{Expected anchors} & \textbf{Observed anchors} & \textbf{Missing} &
\textbf{Step P/R} & \textbf{Chain P/R} & \textbf{\#events} \\
\hline
SC4 &
\texttt{INSTALL, DOWNLOAD, OUTBOUND\_CONN, EXFIL} &
\texttt{INSTALL, DOWNLOAD, OUTBOUND\_CONN} &
\texttt{EXFIL} &
1.00 / 0.75 &
1.00 / 0.75 &
188{,}270 \\
\hline
SC1 &
\texttt{INSTALL, DOWNLOAD, OUTBOUND\_CONN, EXFIL} &
\texttt{INSTALL} &
\texttt{DOWNLOAD, OUTBOUND\_CONN, EXFIL} &
1.00 / 0.25 &
1.00 / 0.25 &
8{,}534 \\
\hline
\end{tabular}
\label{tab:sc_evidence_diag}
\end{table*}

The feasibility matrix (Table~\ref{tab:scenario_method_matrix_binary_no_baseline}) and the above diagnostic patterns suggest that telemetry planning should be driven by \emph{which failure mode dominates} rather than by collecting more logs indiscriminately.

\paragraph{Implication 1: prioritize telemetry that closes missing-phase gaps.}
When failures are dominated by missing-phase gaps, additional correlation logic cannot help: the missing phase must be made observable. Practically, this means adding phase-specific sources (e.g., egress proxy logs for outbound transfer, artifact registry logs for package retrieval, or IAM/API audit for cloud control-plane actions) that expose both the phase signal and its identifiers.

\paragraph{Implication 2: invest in attribution to repair join breaks.}
When steps are present but chains remain fragmented, the bottleneck is attribution. The most effective upgrades are sources or instrumentation that provide stable join keys across layers (process$\leftrightarrow$socket linkage, workload/container identifiers, cloud request IDs). This improves chain continuity without requiring scenario-specific tuning.

\paragraph{Implication 3: treat negative/partial chains as first-class outcomes.}
Attempt signals without downstream anchors should be interpreted as incomplete or failed compromises rather than forced into a full chain. Operationally, requiring downstream confirmation (e.g., attributable outbound sessions or file staging) reduces overestimation of attacker progress and aligns reconstruction with incident response needs.

\paragraph{Implication 4: two-source baselines can be strong but are scenario-dependent.}
A small, complementary set (typically host provenance + network visibility) can be sufficient when it both exposes key phases and provides joinable identifiers. However, scenarios whose critical actions lie outside host/network boundaries (e.g., cloud control-plane misuse or semantic/model-layer attacks) require targeted additional sources; multi-source is therefore most valuable for robustness across heterogeneous scenario structures.

\section{Evidence packages for SC1 and SC4}
\label{app:evidence}

We provide pipeline-generated evidence packages for SC4 (success exemplar) and SC1 (failure exemplar), organized by step anchors
(\texttt{INSTALL}, \texttt{DOWNLOAD}, \texttt{\seqsplit{OUTBOUND\_CONN}}, \texttt{EXFIL}). For each anchor, the package includes (i) timestamped
evidence excerpts with telemetry source attribution, (ii) a step-level time window summary ($t_{\min}$--$t_{\max}$) and evidence volume,
and (iii) single-source vs.\ multi-source ablation indicating which anchors are recoverable from each telemetry source in isolation.

\section{Data Sanitization Details}
\label{app:data-clean}
The collected logs contain environment-specific identifiers that may reveal sensitive information about the deployment. These identifiers include hostnames, non-system user accounts, cloud resource identifiers, and file paths that embed local usernames as substrings. To protect privacy while preserving analytical utility, we apply a stable pseudonymization strategy that replaces such identifiers with consistent tokens across all log sources and scenarios.
We intentionally preserve public Internet indicators to maintain realism in the simulated traffic, while pseudonymizing Azure/Azure-hosted domains that could reveal deployment-specific context.

\subsection{Threat-Model and Design Goals}
The sanitization procedure is designed to satisfy the following goals:
\begin{itemize}
    \item \textbf{Privacy protection:} Obfuscate values that can directly identify infrastructure, users, or internal resources. Well-known system/built-in accounts and placeholder values (e.g., \texttt{SYSTEM}, \texttt{NT AUTHORITY\textbackslash SYSTEM}, \texttt{S-1-...}, \texttt{N/A}) are retained to avoid over-sanitization noise.
    \item \textbf{Consistency across sources:} The same original identifier is mapped to the same pseudonym across all log types and across multiple runs. 
    \item \textbf{Preserve security semantics:} Only identifier fields (e.g., usernames, hostnames, Azure resource IDs, and Azure/Azure-hosted domain names) are pseudonymized; all other fields required for analysis remain unchanged (e.g., event categories, ports, protocols, IP addresses, HTTP methods, DNS query types, and temporal ordering). Public Internet FQDNs are preserved to maintain realism.
    \item \textbf{Determinism:} Sanitization is deterministic under a stable secret salt, enabling reproducible analysis.
\end{itemize}

\subsection{Stable Pseudonymization Mechanism}
We maintain a secret salt value $S$ (stored locally and never published) and use it to derive deterministic tokens. For each sensitive field value $v$, we compute a token
$$
t = \texttt{prefix} || \textsf{H}(S||v),
$$
where $\textsf{H}(\cdot)$ is a cryptographic hash function (e.g., SHA-256) and \texttt{prefix} indicates the identifier category (e.g., \texttt{host\_}, \texttt{user\_}, \texttt{res\_}). In practice, we derive compact tokens (e.g., \texttt{USER\_XYZ}) by hashing $S||v$ (SHA-256), truncating the digest to a numeric identifier, and resolving rare collisions deterministically. We persist a JSON dictionary per category in a shared \texttt{mappings/} directory to ensure stability across files and notebook executions.

This approach preserves equality relationships (same value $\rightarrow$ same token), enabling correlation across sources, without exposing original identifiers. 

\end{document}

%% file: sections/abstract.tex
\begin{abstract}

Advanced software supply chain (SSC) attacks are increasingly runtime-only and leave fragmented evidence across hosts, services, and build/dependency layers, making any single telemetry stream insufficient for chain reconstruction. Despite this, no existing dataset provides multi-source runtime monitoring data with end-to-end chain-level ground truth for SSC attacks, leaving the observability limits of such attacks poorly understood. We present SynthChain, a multi-source runtime dataset with chain-level ground truth derived from real-world malicious packages and exploit campaigns, and use it to empirically study observability limits in SSC attacks. SynthChain covers seven representative SSC exploit scenarios across PyPI, npm, and C++ supply chains, spanning Windows, Linux, and containerized environments, with annotations for 14 MITRE ATT\&CK tactics, 161 techniques, and 2{,}919 manually verified Indicator of Compromise (IOC) annotations across 22 log files spanning 11 telemetry types. Our observability analysis shows that no single source is
chain-complete: even the best single source recovers fewer than 40\% of expected attack steps. Fusing just two complementary sources improves reconstruction by roughly 1.6$\times$, but gains depend on \emph{which} sources are combined rather than how many. We identify three systematic failure modes---missing-phase gaps, attribution breaks, and ambiguity---and derive telemetry planning guidelines that do not require prior knowledge of specific attacks. A preliminary sensitivity analysis confirms that the multi-source advantage persists under reduced per-source sampling rates. The corpus ($\approx$0.59M events) is released with ground truth and artifacts to support reproducible evaluation of runtime SSC defenses.
\end{abstract}

%% file: sections/introduction.tex
\section{Introduction}
\label{sec:introduction}
SSC compromise has become a high-leverage vector for modern adversaries and is ranked among OWASP’s top three threats in 2025~\cite{owasp_top10_2025}. While Mandiant’s 2025 analysis attributes only 0.2\% of initial intrusions directly to supply-chain compromise~\cite{mandianttrends2025}, the downstream blast radius is often disproportionate: recent incidents show how small opportunities can escalate into outsized consequences~\cite{lins2024criticalpathimplantbackdoors,unit42_2025_shaihulud}. Contemporary threat intelligence further indicates that SSC campaigns are no longer isolated package-tampering events, but increasingly manifest as advanced, multi-stage, stealthy operations---including targeted manipulation of AI/ML ecosystems~\cite{reversinglabs_sscs_report}. Echoing this shift, CrowdStrike’s 2025 threat-hunting report highlights cross-domain hands-on-keyboard activity that abuses trusted developer relationships, cloud control planes, and automation~\cite{crowdstrike_threat_hunting_report}.

These trends expose a fundamental gap: \emph{advanced supply-chain attacks unfold across multiple stages and extend beyond source code alone}, and the evidence needed to detect and reconstruct them is \emph{fragmented across heterogeneous telemetry sources, much of it generated at runtime}. Yet existing research and benchmarks predominantly emphasize static artifact inspection (e.g., malicious package discovery)~\cite{malicious-packages,10.1109/ASE56229.2023.00135} or dynamic analysis in limited sandbox settings~\cite{ossf_package_analysis_2024,11025755}. As a result, available datasets offer only partial visibility: static corpora lack execution semantics and post-compromise workflows, while many dynamic datasets rely on relatively simple triggers and do not model the multi-stage, cross-environment distributed behaviors typical of modern supply chain intrusions. Consequently, practitioners and researchers are often forced to study supply chain compromise either without realistic runtime traces or without the multi-source evidence required for chain reconstruction and forensic validation.

\paragraph{Key challenge: observability limits in realistic deployments.}
In real environments, defenders seldom have ``full fidelity'' visibility: telemetry is constrained by access boundaries (e.g., managed services, proprietary build systems), cost and performance budgets, and operational trade-offs (sampling, retention, and source coverage) \cite{nist800137}. For advanced supply-chain intrusions, these constraints are not incidental---they directly determine what is detectable. Evidence is frequently non-redundant across sources: a chain step visible in process lineage can be absent from system logs; a network indicator can be inconclusive without service traces; and pipeline-side artifacts can be inaccessible at runtime \cite{10179405}. This implies that \textbf{single-source detection is inherently incomplete} for advanced supply-chain attacks: even an ideal detector operating on a single stream cannot recover a complete compromise chain when required evidence is missing by design \cite{10.1145/2991079.2991122}.

\paragraph{Our approach.}
Guided by techniques observed in large-scale malicious open-source software (OSS) packages~\cite{malicious-packages}, we select representative real-world samples to construct a near-production supply-chain attack environment and a multi-source runtime dataset. We collect synchronized telemetry across hosts, services, and containers ---including process lineage, system/audit logs, and network/service traces, with container-level visibility enabled via eBPF-based~\footnote{https://ebpf.foundation/} instrumentation ---to support chain-level analysis.

We primarily construct chain-level ground truth tactics, techniques, and procedures (TTPs; the structured vocabulary for describing adversary behavior) by combining (i) technique-level adversary actions directly exported from Mythic\footnote{An open-source C2 framework that records operator tasking and implant responses} C2 tasking logs (ATT\&CK-mapped by construction) with (ii) payload-originated actions extracted via an LLM-assisted pipeline with manual verification. We then align defender-visible events via coarse rule matching.

Finally, We evaluate the marginal benefit of each telemetry stream by comparing single-source, two-source, and multi-source fusion, isolating observability—telemetry anchors and cross-source joins—as the primary limiting factor.

Our key finding is twofold: single-source monitoring is inherently insufficient for chain-complete evidence, while effective multi-source fusion depends on which sources are combined rather than how many. Concretely, our contributions are:
\begin{itemize}
    \item \textbf{Empirical characterization of observability limits in SSC attacks.} 
        We quantify chain-level detectability under realistic telemetry constraints, showing that single-source evidence is inherently incomplete (best single-source recovers $<$40\% of attack steps) and that complementary two-source fusion yields $\approx$1.6$\times$ improvement — but gains depend on \emph{which} sources are fused rather than how many.
    \item \textbf{End-to-end supply-chain scenarios grounded in real incidents.} 
        We distill exploitation patterns from recent incidents into seven representative scenarios spanning PyPI, npm, and native C/C++ supply chains across Windows and Linux, covering all 14 Enterprise ATT\&CK tactics and 161 out of 216 Enterprise techniques (75\% coverage).
    \item \textbf{A multi-source runtime dataset with chain-level ground truth.}
        We release a curated dataset,named SynthChain, with synchronized multi-source telemetry ($\approx$0.59M raw events), 2{,}919 manually verified IOCs annotations across 22 log files, and ATT\&CK-aligned labels, along with the corresponding experimental setups, enabling controlled evaluation of detection systems and forensic reconstruction methods~\cite{10179405,10120960}.
    \item \textbf{Cross-scenario analysis of failure modes and telemetry trade-offs.}
        We identify three systematic failure types — missing-phase gaps, attribution breaks, and ambiguity/noise — and derive actionable telemetry planning guidelines that do not require prior knowledge of specific attacks.
\end{itemize}

To the best of our knowledge, \textbf{SynthChain} is the first public software supply-chain dataset to combine end-to-end multi-stage execution traces, synchronized multi-source telemetry, and manually verified line-level IOC ground truth for chain-level reconstruction under realistic observability constraints.

%% file: sections/motivation.tex
Table~\ref{tab:abbreviations} summarizes the abbreviations used
throughout the paper.

\begin{table}[t]
  \caption{Abbreviations used throughout the paper.}
  \label{tab:abbreviations}
  \centering
  \footnotesize
  \setlength{\tabcolsep}{2.5pt}
  \renewcommand{\arraystretch}{1.0}
  \begin{tabularx}{\columnwidth}{
    @{}
    l >{\raggedright\arraybackslash}X
    l >{\raggedright\arraybackslash}X
    @{}
  }
    \toprule
    \textbf{Abbr.} & \textbf{Meaning}
      & \textbf{Abbr.} & \textbf{Meaning} \\
    \midrule
    SSC   & Software supply chain
          & IOC   & Indicator of compromise \\
    TTP   & Tactics, techniques, and procedures
          & C2    & Command and control \\
    LotL  & Living off the land
          & APT   & Advanced persistent threat \\
    SCA   & Software composition analysis
          & SBOM  & Software bill of materials \\
    StepR & Step recall
          & Recon & Reconstructability \\
    \bottomrule
  \end{tabularx}
\end{table}

\section{Background}
\label{sec:background}
Modern SSC attacks exploit the trust relationships inherent in package ecosystems, dependency resolution, and build/release pipelines \cite{10838587, ohmBackstabberKnifeCollection2020}. Rather than targeting end-user software directly, adversaries compromise upstream components --- such as open-source packages hosted on registries like PyPI and npm --- so that malicious code propagates downstream through legitimate installation and update mechanisms. Entry vectors include typosquatting~\cite{typosquatting} (publishing packages with names resembling popular libraries), dependency confusion~\cite{10179304} (exploiting resolution precedence between public and private registries), and direct compromise of CI/CD credentials or build artifacts. Once a malicious package is installed, the attack typically progresses through multiple post-compromise stages --- payload retrieval, execution, persistence, and data exfiltration --- that may span several hosts, and services~\cite{lazarusgroup, hiddenplainsight}.

To systematically characterize adversary behavior across these stages, we adopt the MITRE ATT\&CK framework,\footnote{https://attack.mitre.org/}, a publicly maintained knowledge base that organizes adversary actions into tactics (the adversary's high-level goals, such as Initial Access, Execution, or Exfiltration) and techniques (the specific methods used to achieve each tactic, such as T1195.002 --- Compromise Software Supply Chain). A technique may have multiple sub-techniques that capture implementation-level variants. Together, extracted ground truth TTPs provide a structured vocabulary for annotating observed behaviors and comparing attack coverage across scenarios. In this work, we map all adversarial actions to ATT\&CK techniques to enable standardized cross-scenario analysis and to support downstream detection evaluation grounded in a shared taxonomy.

A recurring operational pattern in advanced SSC compromises is Living-off-the-Land (LotL), in which adversaries avoid deploying custom tooling and instead abuse legitimate, per-installed system utilities (e.g., PowerShell, curl, or Python interpreters) to execute malicious actions. Because LotL activity produces process and network telemetry indistinguishable from normal administrative workflows, it significantly raises the difficulty of detection from any single telemetry source and motivates the multi-source correlation approach central to our work. Similarly, many of the scenarios we study employ fileless execution—running payloads entirely in memory without writing persistent artifacts to disk—and command-and-control (C2) channels that blend into routine traffic (e.g., HTTPS or SSH sessions), further fragmenting the evidence a defender must piece together across heterogeneous log sources.

\section{Limitations of Existing Telemetry for Advanced Supply-Chain Attack Analysis}
\label{sec:literature_analysis}

Advanced SSC compromises are normally mediated by package ecosystems, dependency resolution, and build/release pipelines, leaving the evidence needed to detect advanced scenarios dispersed across telemetry layers.
We identify two key gaps in current detection practice and research: (i) fragmented observability of multi-stage behaviors across sources, and (ii) a lack of datasets with explicit cross-source alignment to enable chain-level analysis.




\subsection{Fragmented Observability of Advanced Supply-Chain Attacks}
\label{sec:attack_behaviours}

Advanced SSC attacks increasingly minimize localized artifacts by distributing functionality across stages and contexts: trojanised components may rely on LotL to blend into benign activity~\cite{barr-smithExorcistAutomatedDifferential2022}, while Lazarus-attributed incidents illustrate payload fragmentation and encoding across multiple packages to evade static detection~\cite{lazarusgroup}. As a result, evidence is scattered across heterogeneous telemetry (e.g., build/dependency signals, host process activity, and network/service traces) with inconsistent identifiers and loosely synchronized timestamps, making end-to-end reconstruction dependent on explicit cross-source alignment.

Yet most SSC studies and benchmarks remain package-centric, classifying individual packages via static features, ML signatures, or sandboxed traces (e.g., DONAPI~\cite{10.5555/3698900.3699111} and dynamic execution pipelines for npm/PyPI~\cite{10.1145/3691620.3695262}). Even when incorporating inter-package relations (e.g., transitive dependency analysis), linkage is typically established at the code/dependency layer rather than through aligned multi-source runtime evidence~\cite{11121698}; correspondingly, prior surveys largely organize methods around per-instance static/dynamic features~\cite{zhang_malicious_2023}. Overall, SSC defense is thus a chain-level problem spanning dependencies, build infrastructure, and developer-centric workflows~\cite{10.1145/3714464}, motivating datasets and evaluations with chain-level ground truth and explicit cross-source alignment.

\subsection{Synthetic Data Generation and the Lack of Cross-Source Alignment}
\label{sec:data_generation}

To support security evaluation and reproducible experimentation, prior work has proposed synthetic or semi-synthetic datasets and testbeds. One line collects \emph{per-package} behaviors in isolated sandboxes: OpenSSF releases unlabeled execution results with runtime behaviors and static indicators for individual package instances~\cite{ossf_package_analysis_2024}, and QUT-DV25 provides large-scale dynamic traces for PyPI SSC attacks using eBPF-based kernel and user-level probes~\cite{mehedi2025qutdv}. While valuable for package-level detection, these resources typically treat each package as the unit of analysis and lack explicit cross-source alignment or chain-level ground truth across stages.

Another line builds simulation-based testbeds, e.g., model-driven environments for infrastructure and attack behaviors~\cite{landauer_have_2021} with improved realism via user-activity simulation~\cite{landauer_maintainable_2023}, but they often focus on limited telemetry (mainly system logs and network traffic) and do not model supply-chain–specific propagation paths. Large-scale semi-synthetic Advanced Persistent Threat (APT) datasets demonstrate multi-stage trace generation~\cite{edq8-nk52-21,MYNENI2023109688}, yet they rely on restricted telemetry, do not explicitly encode cross-layer alignment, and capture general APT behaviors rather than supply-chain executions governed by dependency resolution and package-driven propagation~\cite{10838587}. Other synthetic corpora emphasize traffic diversity, attack variety, or labeling quality~\cite{Sharafaldin2018TowardGA,10286332,s23135941}, which suits IDS benchmarking but not stealthy SSC chains.

Overall, existing data generation efforts emphasize realism or scale, but seldom address the \emph{cross-source alignment} needed for chain reconstruction, where semantically related events must be correlated across heterogeneous telemetry.

\section{Related Work}

To inform our unique experimental setup and select representative scenarios that reflect recent SSC exploitation trends, we compare our dataset against prior datasets in terms of covered telemetry sources. We also perform a statistical analysis of a large corpus of malicious open-source packages, using their documented malicious functions and behaviors, to characterize technique usage trends and guide scenario selection.

\subsection{Dataset Comparison}

Table~\ref{tab:rw_capability} compares representative datasets and testbeds against capabilities required for chain-level supply-chain analysis, distinguishing package-level corpora~(a) from end-to-end attack scenario datasets~(b). {Multi-Stage and Multi-Source indicate whether a dataset captures full attack progressions and heterogeneous telemetry, respectively; IOC ground truth (GT) denotes the availability of manually verified, line-level IOC annotations with source attribution, which is essential for deterministic chain reconstruction and evaluating correlation or provenance reasoning. Related provenance-based detection and forensic investigation methods reason over available system evidence~\cite{cai2026building,Cheng2023KairosPI,10179405,NODLINK2024,10120960}, whereas we study whether evidence remains observable and joinable across heterogeneous supply-chain stages and telemetry boundaries. We additionally report on ATT\&CK mapping, Tracee/eBPF host tracing, and the presence of normal background behavior.

\begin{table*}[t]
\centering
\footnotesize
\caption{Capabilities of existing datasets/testbeds vs.\ requirements for chain-level supply-chain attack analysis}
\label{tab:rw_capability}

\vspace{1mm}
\noindent\textbf{(a) Package-level corpora.} Each entry represents an isolated single-package analysis instance; counts reflect individual packages, not end-to-end attack scenarios.

\vspace{1mm}
\begin{tabular}{@{}lcccccccc@{}}
\toprule
\textbf{Work} &
\textbf{Scale} &
\textbf{SC} &
\textbf{Multi-Stage} &
\textbf{Multi-Source} &
\textbf{IOC GT} &
\textbf{ATT\&CK TTPs} &
\textbf{Tracee/eBPF$^{*}$} &
\textbf{Normal Behavior} \\
\midrule
QUT-DV25 (2025) \cite{mehedi2025qutdv} & $\sim$14K pkgs & \checkmark & -- & \checkmark & -- & -- & \checkmark & -- \\
OpenSSF (2025) \cite{ossf_package_analysis_2024} & $\sim$16K pkgs & \checkmark & -- & \checkmark & -- & -- & -- & -- \\
Zhang et al.\ (2025) \cite{zhang_malicious_2023} & $\sim$10K pkgs & \checkmark & -- & -- & -- & -- & -- & -- \\
Backstabber (2020) \cite{ohmBackstabberKnifeCollection2020} & 174 pkgs & \checkmark & -- & -- & -- & -- & -- & -- \\
\bottomrule
\end{tabular}
\label{tab:rw_package}

\vspace{3mm}
\noindent\textbf{(b) End-to-end attack scenario datasets.} Each entry models complete multi-stage compromise chains; counts reflect distinct attack scenarios with runtime telemetry.

\vspace{1mm}
\begin{tabular}{@{}lcccccccc@{}}
\toprule
\textbf{Work} &
\textbf{\#Scenarios} &
\textbf{SC} &
\textbf{Multi-Stage} &
\textbf{Multi-Source} &
\textbf{IOC GT} &
\textbf{ATT\&CK TTPs} &
\textbf{Tracee/eBPF$^{*}$} &
\textbf{Normal Behavior} \\
\midrule
Windows-APT (2025) \cite{MOZAFFARI2026112569} & 36$^{\dagger}$ & -- & \checkmark & -- & $\sim$ & \checkmark & -- & \checkmark \\
Linux-APT (2024) \cite{KARIM2024110290} & 5 & -- & \checkmark & \checkmark & $\sim$ & \checkmark & -- & \checkmark \\
Landauer et al.\ (2023) \cite{landauer_maintainable_2023} & 8 & -- & \checkmark & \checkmark & -- & \checkmark & -- & \checkmark \\
Unraveled (2023) \cite{MYNENI2023109688} & 3 & -- & \checkmark & \checkmark & -- & \checkmark & -- & \checkmark \\
OpTC (2021) \cite{edq8-nk52-21} & 5 & -- & \checkmark & \checkmark & -- & -- & -- & \checkmark \\
\textbf{SynthChain (this work)} & \textbf{7} & \checkmark & \checkmark & \checkmark & \checkmark & \checkmark & \checkmark & \checkmark \\
\bottomrule
\end{tabular}
\label{tab:rw_scenario}

\vspace{2mm}
\noindent
\textbf{SC}: supply-chain specific;
\textbf{IOC GT}: line-level IOC annotations with source attribution enabling deterministic chain reconstruction;
$\sim$: partial (automated technique-level logs from emulation framework, not manually verified line-level annotations);
$^{*}$eBPF-based host tracing (e.g., Tracee);
$^{\dagger}$Windows-APT 2025 uses Caldera automated emulation; scenarios are technique-sequence playbooks rather than end-to-end attack chains with realistic payload execution.
\end{table*}

Package-level corpora provide large-scale coverage of individual malicious packages but treat each package as an isolated analysis instance, lacking multi-stage traces, cross-source alignment, or chain-level ground truth. End-to-end scenario datasets offer multi-stage, multi-source traces; however, existing testbeds are not supply-chain specific and typically provide only coarse ground truth---narrative-level red-team reports~\cite{edq8-nk52-21}, binary benign/malicious labels~\cite{MYNENI2023109688,landauer_maintainable_2023}, or automated emulation logs~\cite{MOZAFFARI2026112569,KARIM2024110290}---rather than line-level IOC annotations tied to specific log files and records. To our knowledge, SynthChain is the first to combine supply-chain--specific scenarios with multi-stage, multi-source telemetry, manually verified line-level IOC ground truth (2{,}919 annotated records across 22 log files), ATT\&CK-grounded TTPs, eBPF-based tracing, and realistic background activity.

\subsection{Scenario Selection and Representativeness}
We consider seven scenarios: Stegano and Starter model PyPI typosquatting; Parallel and NPMEX model npm lifecycle/dependency attacks; 3CX represents trojanized software; CloudEX models CI/CD credential and artifact compromise; and LayerInj represents a backdoored ML model.
To validate that our scenario selection is representative, we assess coverage at two complementary levels:
(i)~package-level behavioral categories observed in the OpenSSF corpus~\cite{ossf_package_analysis_2024},
and (ii)~supply-chain attack vectors defined in the Ladisa et al.\ SoK taxonomy~\cite{10179304}.

\textit{Package-level coverage.}
We map the seven scenarios against the behavioral distributions extracted from the 16{,}272-package OpenSSF corpus (Figure~\ref{fig:pkg-taxonomy} in Appendix~\ref{app:malicious package statistic analysis table}). For \emph{triggers}, our scenarios collectively cover installation-time
and download-time activation, which together account for
99.7\% of all packages with a trigger label (15{,}583 and 594 of 16{,}223 labeled instances, respectively).
For \emph{malicious functions and objectives}, all seven scenarios include at least one of
payload delivery, data exfiltration, or credential/data theft---the three
most frequent function categories, which jointly appear in over 90\%
of labeled packages. 
For \emph{evasion methods}, our scenarios span encoding/obfuscation
(covering 98.7\% of evasion-labeled packages, i.e., 6{,}360 of 6{,}443)
and additionally include steganography
from the long tail (0.4\%). 

\textit{Beyond package-level behaviors.}
Crucially, the OpenSSF corpus captures only \emph{per-package} malicious behaviors observed in isolated sandbox executions. Our end-to-end scenarios instantiate substantially richer post-compromise techniques that are invisible at the package level, including file replacement and persistence via startup modification (SC2),
DLL side-loading and process injection (SC5), CI/CD credential abuse and artifact tampering (SC6), and backdoored ML model deployment with conditional triggers (SC7)~\cite{reversinglabs_sscs_report,crowdstrike_threat_hunting_report}.
These techniques emerge only when attack chains extend beyond the package boundary into host, network, and cloud/container environments---precisely the multi-stage
behaviors that package-level corpora cannot capture.

\textit{Alignment with SoK attack taxonomy.}
Mapping our scenarios to the supply-chain attack taxonomy of Ladisa et al.~\cite{10179304}, SynthChain covers three of the four top-level attack vectors: \emph{name confusion} (typosquatting in SC1, SC2), \emph{subverting legitimate packages} (dependency-chain manipulation in SC3, SC4; trojanized software in SC5; CI/CD compromise in SC6), and \emph{developing and advertising distinct malicious packages} (malicious ML model artifacts in SC7).
The fourth vector---\emph{compromising the package repository infrastructure itself}---falls outside our scope, as it targets registry-level mechanisms
(e.g., account takeover of maintainers or manipulation of registry metadata)
that do not produce the host- and network-level runtime telemetry
that is the focus of this work.

\begin{table*}[t]
\centering
\caption{Scenario Coverage Matrix}
\label{tab:scenario-coverage}
\scriptsize
\setlength{\tabcolsep}{3.5pt}
\renewcommand{\arraystretch}{1.15}

\begin{tabular*}{\textwidth}{@{\extracolsep{\fill}}@{}l *{17}{c}@{}}
\toprule
& \multicolumn{5}{c}{Trigger} & \multicolumn{7}{c}{Evasion} & \multicolumn{5}{c}{Functions} \\
\cmidrule(lr){2-6}\cmidrule(lr){7-13}\cmidrule(lr){14-18}
Case
& Inst & DL & Hook & CICD & Cond
& Obf & Steg & Enc & FRep & MS & Inj & Fileless
& Exfil & C2 & Steal & Payload & Persist \\
\midrule
1.Stegano
& \cmark & \cmark &  &  &
& \cmark & \cmark & \cmark &  &  &  & \cmark
& \cmark & \cmark & \cmark & \cmark &  \\
2.Starter
& \cmark & \cmark  &  &  &
& \cmark &  & \cmark & \cmark & \cmark &  & \cmark
& \cmark & \cmark & \cmark & \cmark & \cmark \\
3.Parallel
& \cmark & \cmark & \cmark &  &
& \cmark &  & \cmark &  & \cmark &  & \cmark
& \cmark & \cmark & \cmark & \cmark &  \\
4.NPMEX
& \cmark & \cmark & \cmark &  &
& \cmark &  & \cmark &  & \cmark &  &
& \cmark &  & \cmark & \cmark &  \\
5.3CX
& \cmark & \cmark &  &  &
& \cmark &  & \cmark &  &  & \cmark &
&  & \cmark & \cmark & \cmark & \cmark \\
6.CloudEX
& \cmark &  &  & \cmark & 
& \cmark &  & \cmark  &  &  &  & \cmark
& \cmark & \cmark & \cmark & \cmark &  \\
7.LayerInj
& \cmark &  &  &  & \cmark
& \cmark &  & \cmark &  &  &  & \cmark
& \cmark & \cmark & \cmark & \cmark & \cmark \\
\bottomrule
\end{tabular*}

\vspace{2pt}
\footnotesize{\textit{Legend: CICD=CI/CD pipelines; Cond=conditional trigger; Enc=encoding/stream cipher;
FRep=file replacement; MS=multi-stage/sequenced execution; Inj=DLL side-loading/process injection; Fileless=fileless malware;
Steal=local data theft; Payload=payload download.}}
\end{table*}

Table~\ref{tab:scenario-coverage} confirms per-scenario coverage: every dominant behavioral category is exercised by at least two scenarios, while long-tail techniques (e.g., steganography, conditional triggers) are each instantiated by at least one scenario to preserve diversity.
In summary, although the scenario count is seven---comparable to or exceeding other end-to-end attack testbeds or datasets (Table~\ref{tab:rw_capability}b)---the \emph{behavioral coverage} spans the categories that collectively characterize $>$\textbf{95\%} of observed malicious packages in the wild.

%% file: sections/methodology.tex
\section{Methodology}
\label{sec:methodology}

This section details how we construct the SynthChain dataset. To enable controlled, reproducible collection of multi-source runtime telemetry across diverse SSC attack scenarios, we develop a structured emulation and analysis framework (Figure~\ref{fig: Infrastructure}). This framework emulates end-to-end, multi-stage compromise pathways in controlled environments and prioritizes system-level observability over implementation details. We retain full-stage behaviors up to exfiltration to support early-stage detection, while omitting generic reconnaissance that is not characteristic of typical supply-chain exploitation. The methodology covers system setup, telemetry collection, monitoring configuration, benign-behavior emulation, design principles, and the resulting attack scenarios.

\subsection{Setting Up}
\label{sec:setting_up}
Our testbed approximates realistic development, deployment, and cloud-integrated supply-chain environments. It includes Windows and Linux hosts, as well as Docker-based workloads to emulate AI-component integrations common in modern pipelines.

\begin{figure*}[htbp]
    \centering
    \footnotesize
    \includegraphics[scale=0.6]{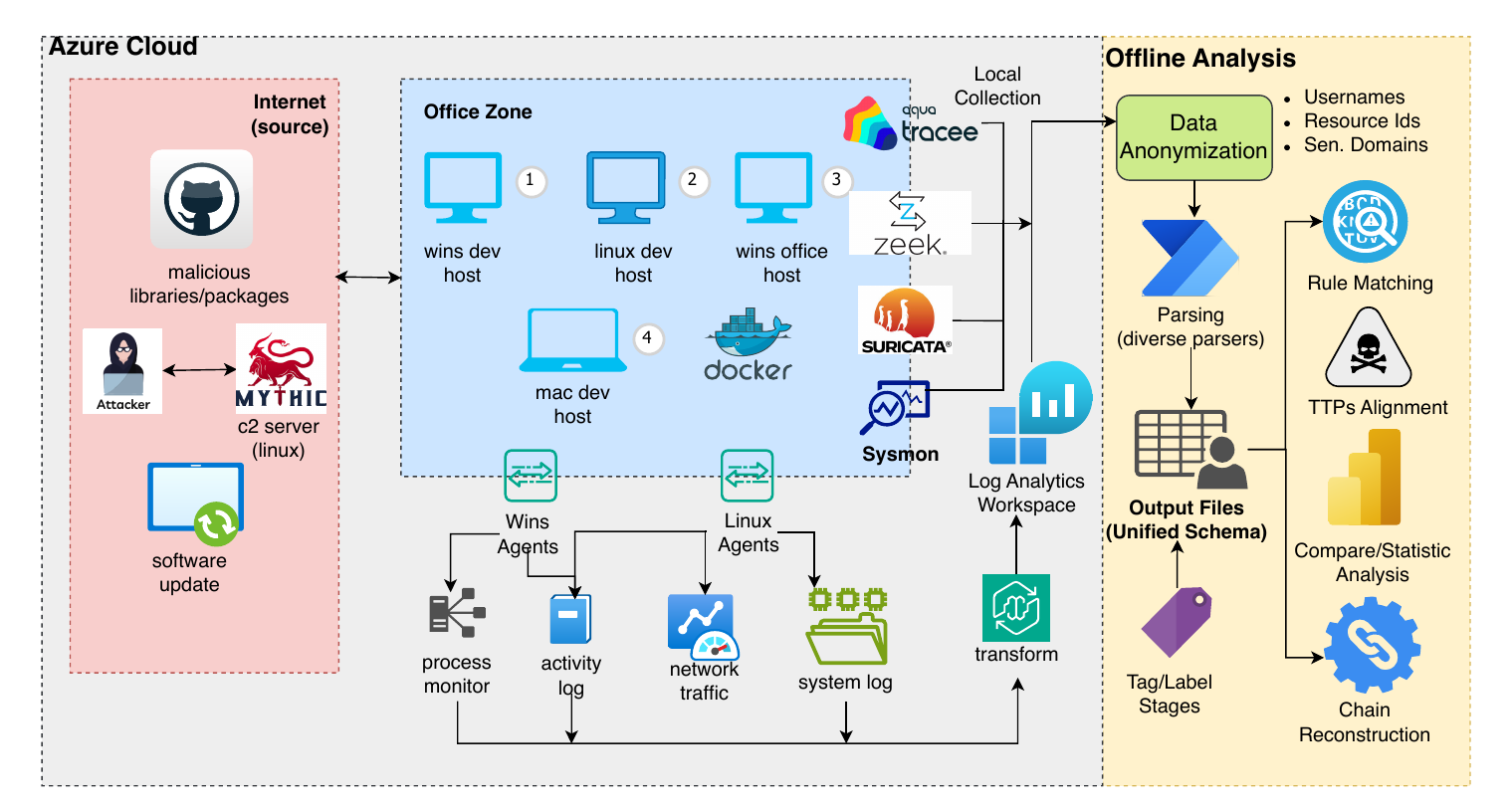}
    \caption{Simulation Workflow and Analysis Pipeline}
    \label{fig: Infrastructure}
\end{figure*}

Telemetry is collected via two ingestion paths and then processed by a common post-processing pipeline, as demonstrated in Figure~\ref{fig: Infrastructure}. For sources natively supported by Azure Log Analytics (e.g., Windows events and Syslog), logs are ingested into the workspace and passed through a lightweight transformation layer to normalize schemas and fields. For other sources (e.g., Zeek and Suricata), we directly extract records from hosts and feed them into the same normalization stage. All streams then undergo common parsing and stable salted
pseudonymization of deployment-specific identifiers, preserving
cross-source equality relationships and security-relevant semantics;
the full sanitization procedure and threat model are provided in
Appendix~\ref{app:data-clean}. The environment contains attacker-controlled infrastructure, development hosts, office hosts, and public-internet access that supports download and update activities. 

\subsubsection{Collected Data}
SynthChain integrates telemetry from heterogeneous environments, including Windows hosts, Linux hosts, and Docker-based container workloads. We collect host-, network-, and system/authentication logs, and optionally enrich them with behavioral tracing for higher-fidelity action reconstruction. Detailed data sources by platform are summarized in provided artifacts.

\paragraph{Telemetry variability and trust domains.}
Telemetry availability and granularity vary across scenarios due to differences in execution environments and attack outcomes (e.g., partial execution, fileless activity, and container-scoped behaviors). Rather than enforcing artificial completeness, we preserve these natural observability gaps to support analysis of when single-source telemetry fails and how multi-source evidence mitigates such limitations.

When target-side telemetry is sparse, attacker-side and C2/operator logs are retained as auxiliary provenance and used for ground-truth and ATT\&CK annotation. Each record carries a collection-origin tag, and all attacker-side records are excluded before step tagging, event-graph construction, chain reconstruction, and computation of every reported observability metric.

\paragraph{MITRE ATT\&CK-Aligned Data Annotation}

To ensure consistent and interpretable labeling across heterogeneous traces, we annotate all adversarial actions using MITRE ATT\&CK techniques.
Our annotation follows two complementary paths depending on the action's origin: (i) \emph{C2-driven actions}, whose technique mappings are directly exported from the Mythic\footnote{
https://docs.mythic-c2.net/home} framework's built-in ATT\&CK
tagging of recorded operator tasks; and (ii) \emph{payload-originated
actions}, whose mappings are produced by an LLM-assisted pipeline with subsequent human validation, detailed in Appendix~\ref{app:llm-assisted analysis}. Together, these
annotations form the semantic backbone for scenario construction
and subsequent threat analysis, and also enable cross-scenario
technique distribution analysis (Appendix~\ref{app:ttp-distribution}).

\subsubsection{Normal Behavior Modeling}

To elicit realistic runtime signals, we inject benign background activity that commonly co-occurs with early-stage supply-chain compromises (Appendix~\ref{app:benign-activity}, Table~\ref{tab:behaviour_definition}). Unlike prior work using predefined attack scripts or controlled workloads \cite{landauer_maintainable_2023, edq8-nk52-21}, our environment embeds routine usage patterns that introduce realistic noise and may obscure stealthy exploitation. Activities include downloads and updates, filesystem operations, outbound web communication, and interactive use (e.g., browsing, office work, and service execution).

To increase diversity, hosts are assigned functional profiles (development vs.\ office), yielding different process types, and communication patterns, which makes rare anomalies harder to isolate and better reflects real detection conditions. We do not model human intent; instead, we randomize and schedule sufficient benign variability for meaningful forensic analysis (Appendix~\ref{app:benign-activity}).

\subsection{Design Principles}
\label{sec:design_principles}

Our scenarios are designed to \textbf{balance realism, representative supply-chain threat coverage, and resistance to trivial detection}. Below, we highlight the key principles that guide the construction of our attack behavior. 

\paragraph{Randomness}
Randomness plays a dual role in our scenario design. We use randomness to avoid overly regular traces. For benign activity, randomized scheduling and mixed activity types emulate natural operational irregularity and provide realistic background noise. For adversarial behavior, we randomize triggers, command ordering, timing, and delivery paths to prevent fixed workflows that would otherwise yield easy signatures.

\paragraph{From specific incidents to reusable attack patterns.}
Our scenarios are not synthetic in the sense of being artificially simplified or stripped of specificity.  Each scenario is grounded in a documented real-world incident  (Section~\ref{sec:scenarios}) and preserves its characteristic techniques---e.g., LSB steganography in SC1~\cite{hiddenplainsight}, DLL side-loading in SC5~\cite{3CX}, and CI/CD credential 
abuse in SC6~\cite{cloudattack}. What we abstract away are environment-specific identifiers (hostnames, credentials, internal URLs) that would limit 
reproducibility, not the attack semantics themselves. The term \emph{synthetic} refers to the controlled re-execution of real attack logic in an instrumented testbed---analogous to how DARPA TC~\cite{edq8-nk52-21} and Unraveled~\cite{MYNENI2023109688} replay real APT campaigns in laboratory settings---not to the generation of artificial or simplified attack patterns.

\paragraph{Adversarial Goals}
We model APT-like SSC adversaries focused on covert information theft and persistence. Scenarios stress operational security: lightweight obfuscation/encoding to frustrate superficial inspection, staged execution with minimal observable footprint, and (when applicable) in-memory execution to reduce disk artifacts and hinder file-centric defenses and forensics. Exfiltration is modeled as selective and low-noise to reflect realistic theft-oriented behavior.

\paragraph{Comparable End-to-End Attack Semantics}

To support systematic comparison across scenarios, each attack chain ends with an explicit exfiltration phase. If a sample already implements exfiltration, we preserve it; otherwise, we only add minimal external orchestration to complete missing stages without altering the intended semantics.

\subsection{Scenarios}
\label{sec:scenarios}

Based on our environment and telemetry pipeline, we construct controlled SSC attack scenarios that capture end-to-end multi-stage behaviors across heterogeneous environments, focusing on host-level observability. Each scenario is derived from a documented \textbf{real-world incident or campaign} (using real malicious packages where available, i.e.\ SC1--SC5), and collectively the seven scenarios cover the dominant trigger, evasion, and objective categories identified in our statistical analysis of 16{,}272 malicious packages. Scenario designs (triggers/evasion, key functions, and tools) are summarized in Table~\ref{tab:attack scenarios} (Appendix~\ref{app:attack-flows}).

All evidence is derived solely from system telemetry (e.g., process creation, file I/O, network connections, and package manager activity). If a sample lacks a native C2/exfiltration mechanism, we use Mythic as a controlled C2 endpoint to complete the chain; otherwise, we preserve the sample's original behavior. Mythic agents (Apollo, Medusa) cover both script-based and compiled payload delivery. For heavily packed samples (e.g., 3CX), we rely on runtime observables rather than unpacking.

\subsubsection{SC1 (Stegano) and SC2 (Starter)}

SC1 and SC2 are both Windows typosquatting cases derived from the Checkmarx campaign report~\cite{hiddenplainsight}, SC1 leverages LSB steganography to conceal a payload inside an image retrieved at install time, followed by in-memory execution and C2-based exfiltration. SC2 instead implements an explicit multi-stage chain emphasizing persistence through startup-folder modification, with staged payload retrieval and execution(workflow in Appendix~\ref{app:attack-flows} with Figure~\ref{fig:attack scenario 1 and 2}).

\subsubsection{SC3 (Parallel) and SC4 (NPMEX)}

SC3 and SC4 target the Linux/npm ecosystem. SC3, based on reported npm incidents~\cite{Phylum, githubfilelessmalare}, models lifecycle-hook-triggered parallel script execution with detached reconnaissance and exfiltration. SC4, inspired by Lazarus-attributed dependency-chain attacks~\cite{lazarusgroup}, captures sequential multi-package execution where artifacts are exchanged between dependencies before dynamic code retrieval and payload deployment( workflow in Appendix~\ref{app:attack-flows} with Figure~\ref{fig:attack scenario 3} and Figure~\ref{fig:attack scenario 4}).

\subsubsection{SC5 (3CX)}

Based on public reports of the 3CX incident \cite{3CX}, we simulate a simplified chain comprising trojanised installer execution, DLL side-loading, in-memory payload execution, and subsequent C2 attempts.
To preserve semantic fidelity, we execute collected samples in their original binary form; execution is conducted in a constrained environment for safety and observability (workflow in Appendix~\ref{app:attack-flows} with Figure~\ref{fig:attack scenario 5}).

\subsubsection{SC6 (CloudEX)}

CloudEX models cloud-based supply-chain compromise targeting CI/CD pipelines and build artifacts, adapted from reported cases \cite{cloudattack}.
We abstract four stages: (1) initial access to an exposed service; (2) discovery of residual CI/CD credentials; (3) access to internal artifact repositories; and (4) artifact modification to propagate downstream impact (workflow in Appendix~\ref{app:attack-flows} with Figure~\ref{fig:attack scenario 6}).

\subsubsection{SC7 (LayerInj)}

LayerInj models ML supply-chain attacks where a tampered model artifact embeds a persistent backdoor without explicit malicious code \cite{10.1109/ICSE43902.2021.00035, wickens2024shadowlogic,wunderwuzzi2024backdooring}.
We abstract three stages: (1) introduction of a tampered model; (2) deployment in a downstream service; and (3) trigger-based activation at inference time (workflow in Appendix~\ref{app:attack-flows} with Figure~\ref{fig:attack scenario 7}).


%% file: sections/analysis.tex
\section{Attack Scenario Analysis}
\label{sec:scenario_analysis}

This section analyzes how the defined supply-chain attack scenarios manifest in observable multi-source telemetry, abstracting away attack implementation details and focusing exclusively on execution traces and extracted indicators.
Our analysis is organized around three system-level questions:
\begin{enumerate}
    \item \textbf{Q1}: How completely can end-to-end attack chains be reconstructed from the available telemetry?
    \item \textbf{Q2}: Where and why does single-source telemetry fail to support reliable reconstruction?
    \item \textbf{Q3}: How does multi-source telemetry mitigate these failures across different attack structures?
\end{enumerate}

The answers to these questions form the basis for our cross-scenario insights and deployment implications.

\subsection{Analysis Methodology}
\label{sec:scenario_method}

We treat defender-visible victim-side telemetry as the only evidence base and apply a uniform pipeline that (i) parses and normalizes heterogeneous logs into a schema-tolerant event table, (ii) tags events with coarse behavioral steps, and (iii) reconstructs candidate attack chains by correlating evidence across time and entities.

\paragraph{(1) Scenario-scoped ingestion and normalization.}
Parsers convert raw records into a unified table with a canonical timestamp field ($ts$) and a lightweight text blob (e.g., \texttt{raw}/\texttt{message}) for matching; $ts$ is normalized to a consistent time basis for stable ordering.

\paragraph{(2) Coarse step tagging (intermediate evidence).}
To align heterogeneous telemetry under an operational defender model, the tagger operates only on collected runtime records, rather than payload source code, C2 tasking logs, or scenario scripts, which are used only offline for ground-truth specification and evaluation. 
We therefore use five telemetry-observable phase anchors—\texttt{INSTALL}, \texttt{AUTH}, \texttt{DOWNLOAD}, \texttt{OUTBOUND\allowbreak\_{}{\allowbreak}CONN}, \texttt{EXFIL}—as a portable reconstruction layer rather than as replacements for fine-grained ATT\&CK techniques. Their selection abstracts recurring, telemetry-visible stages identified across our scenario and representativeness analyses (Table~\ref{tab:scenario-coverage}). We intentionally favor conservative, schema-tolerant rules over recall-optimized tagging so that anchor assignments remain reproducible, auditable, and manually verifiable across telemetry sources. ATT\&CK mappings remain available for technique-level interpretation but are not used as per-event technique detectors.

\begin{table*}[htbp]
\centering
\caption{Behavioral anchors and representative MITRE ATT\&CK techniques}
\label{tab:behavioral-anchors}
\small
\begin{tabularx}{\textwidth}{@{}l X l X@{}}
\toprule
\textbf{Anchor} &
\textbf{Definition and rationale} &
\textbf{Representative ATT\&CK} &
\textbf{Ambiguity / finer alternative} \\
\midrule

\texttt{INSTALL} &
Package, dependency, or lifecycle activation; common SSC entry/trigger &
T1195.002, T1059 &
Benign installs/updates; separate resolution, installation, and hook execution. \\

\texttt{AUTH} &
Credential, token, or service authentication/use &
T1078, T1528 &
Legitimate login/token refresh; separate credential access, credential use, and API authentication. \\

\texttt{DOWNLOAD} &
Retrieval of a dependency, artifact, or subsequent payload &
T1105 &
Benign dependency fetch/update; separate dependency, artifact, and second-stage retrieval. \\

\texttt{OUTBOUND\_CONN} &
Externally directed network session associated with the chain &
T1071, T1090 &
Routine web/API traffic; separate DNS, connection, TLS, and confirmed C2. \\

\texttt{EXFIL} &
Outward transfer consistent with data or artifact movement &
T1041, T1020, T1030 &
Legitimate upload/backup; separate collection, staging, and confirmed transfer. \\

\bottomrule
\end{tabularx}
\end{table*}

\paragraph{ (2b) IOC-based ground truth labeling.}

For each scenario we perform line-level Indicator of Compromise (IOC) manual annotation (e.g., traffic patterns, package names, IP address, ports, and other static artifacts) across every log file to establish ground truth labels (Table~\ref{tab:ioc-gt}). Across 22 log files spanning 11 telemetry types, we identify \textbf{2{,}919 IOC records} in total, distributed across exactly half (22) of the files; the remaining 18---including \texttt{auth.log}, \texttt{zeek\_{}ssh}, and \texttt{suricata.log}, etc---contain no attack traces, reflecting the realistic sparsity of attack signals in multi-source environments. Per-scenario IOC counts range from 51 (sc4) to 833 (sc5). The distribution reveals a clear platform-dependent pattern: Windows scenarios concentrate IOCs in Sysmon-based \texttt{azure\_events}, which accounts for 1{,}761~records (60.3\% of all IOCs),, with minor contributions from \texttt{azure\_conn} and \texttt{azure\_process}. 
In contrast, Linux scenarios distribute traces across Suricata's \texttt{eve.json} (420~records, 14.4\%) and multiple Zeek network logs. The specific Zeek sources activated depend on attacker protocol choice: \texttt{zeek\_ssl} captures IOCs only in sc4 where the C2 uses HTTPS with a self-signed certificate, while \texttt{zeek\_http} records callbacks in sc3 and sc7 where plaintext HTTP is used. Each record is annotated with its exact source file and line number(s), enabling deterministic reproducibility and per-stage evaluation of detection systems. Further IOC coverage and data quality discussion are covered in Appendix~\ref{app:attack-flows}.

\begin{table}[t]
\centering
\caption{IOC ground truth distribution. \textbf{\#F}: log files collected; \textbf{\#IOC}: unique labeled lines. Distribution lists files with $\geq$1 IOC (count); files with none are omitted.}
\label{tab:ioc-gt}
\small
\setlength{\tabcolsep}{2.5pt}
\begin{tabular}{
  @{}
  c
  l
  @{\hspace{1.5pt}}
  r
  @{\hspace{1.5pt}}
  r
  @{\hspace{3pt}}
  l
  @{}}
\toprule
\textbf{ID} & \textbf{Malicious Software} & \textbf{\#F} & \textbf{\#IOC} & \textbf{Distribution across Logs} \\
\midrule
sc1 & \texttt{colorsapi-6.6.7}          &  5 &   656 & \textsf{AE}(473) \textsf{AC}(141) \textsf{AP}(42) \\
sc2 & \texttt{pystallerer-1.0.0}        &  1 &    78 & \textsf{AE}(78) \\
sc3 & \texttt{olymptrade}               & 11 &   289 & \textsf{EJ}(170) \textsf{ZC}(96) \textsf{AS}(13) \textsf{ZD}(6) \textsf{ZH}(4) \\
sc4 & \texttt{audit-ejs|-vue}           & 11 &    51 & \textsf{AS}(26) \textsf{EJ}(12) \textsf{ZC}(6) \textsf{ZD}(4) \textsf{ZS}(3) \\
sc5 & X\_TRADER (3CX)                   &  1 &   833 & \textsf{AE}(833) \\
sc6 & Cred.\ leak$\to$\texttt{setup.py} &  1 &   377 & \textsf{AE}(377) \\
sc7 & Malicious ML model                & 10 &   635 & \textsf{EJ}(238) \textsf{ZC}(155) \textsf{AS}(84) \textsf{ZH/ZF}(77) \textsf{ZD}(4) \\
\midrule
    & \textbf{Total}                     & \textbf{40} & \textbf{2{,}919} & 22/40 files contain IOCs \\
\bottomrule
\end{tabular}
\vspace{2pt}
{\scriptsize
\textsf{AE}\,\texttt{azure\_events}\;
\textsf{AC}\,\texttt{azure\_conn}\;
\textsf{AP}\,\texttt{azure\_process}\;
\textsf{EJ}\,\texttt{eve.json}\;
\textsf{AS}\,\texttt{azure\_syslog}\;
\textsf{ZC}\,\texttt{zeek\_conn}\;
\textsf{ZD}\,\texttt{zeek\_dns}\;
\textsf{ZF}\,\texttt{zeek\_files}\;
\textsf{ZH}\,\texttt{zeek\_http}\;
\textsf{ZS}\,\texttt{zeek\_ssl}
}
\end{table}

\paragraph{(3) Event correlation \& chain reconstruction.}
We propose a two-stage approach: first constructing a fine-grained temporal event graph, then extracting ordered attack chains over it. 

\textbf{Stage1: Temporal event graph construction}. Given tagged events,
we build a directed graph $G$ whose nodes are event records and whose
edges represent observable-dependency proxies rather than definitive
causality. Events within the sliding window $W$ are linked using four
correlation cues: host-local temporal adjacency; shared host entities
(\texttt{pid}, \texttt{ppid}, or user); shared cross-host flow/session
identifiers (\texttt{conn\_id}, \texttt{uid}, or \texttt{flow\_id}); and
shared source--destination IP pairs when explicit session identifiers are
unavailable.
Each stage is annotated with its correlation type (used as label) and $\Delta t$, allowing downstream analysis to distinguish stronger continuity cues (e.g., shared process or flow identifiers) from weaker association cues
(e.g., IP-pair or temporal adjacency).

\textbf{Stage2: Chain reconstruction}. 
For each attack scenario, we use the scenario-specific partial order defined in step tagging to identify anchor events, i.e., the earliest event matching each expected attack step, and attempt to connect adjacent anchors via shortest paths in $G$.
Each inter-step link is assigned one of three statuses:

\begin{itemize}
    \item \textbf{Connected.} A directed path of at most $H$ hops (default $H=8$) exists in~$G$ between two consecutive anchors, indicating a fully traceable causal chain through observable events.
    \item \textbf{Weakly connected.} No causal path exists in-$G$, but the later anchor occurs after the earlier one within a temporal bridge window of (default $B{=}3{,}600$). This indicates temporal plausibility rather than observable causality.
    \item \textbf{Gap (breakpoint).} Neither a causal path nor a plausible temporal bridge exists, indicating a hard discontinuity in the reconstructed chain.
\end{itemize}

\begin{figure*}[t]
    \centering
    \footnotesize
    \includegraphics[width=1\linewidth]{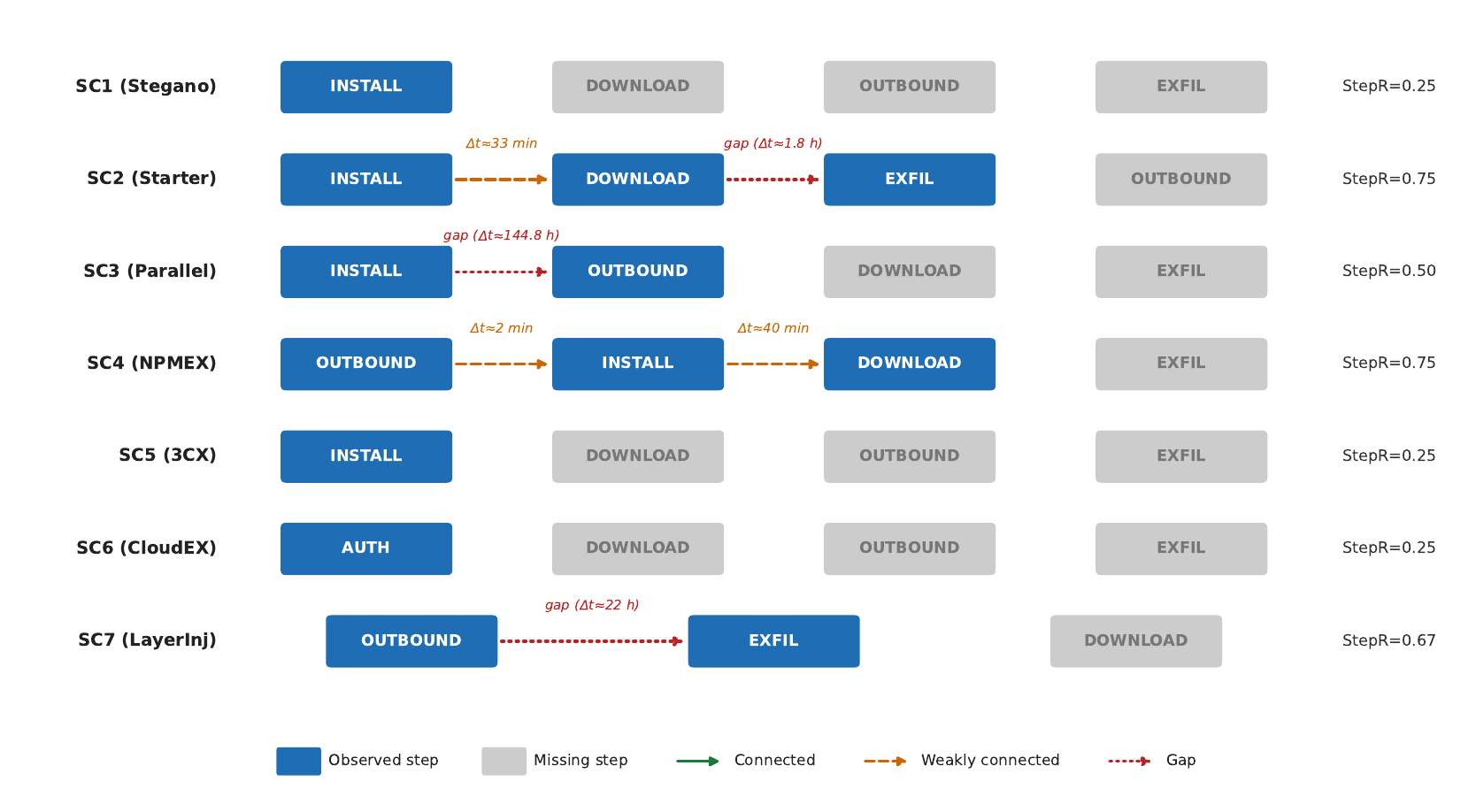}
    \caption{Chain Reconstruction Results Across Scenarios}
    \label{fig:chain_recon}
\end{figure*}

The output is an ordered sequence of (step, anchor, link-status, $\Delta t$) tuples that constitutes the candidate attack chain. Figure~\ref{fig:chain_recon} illustrates the reconstruction results across all seven scenarios. Across all inter-step links, none achieved \emph{connected} status: four were classified as \emph{weakly connected} (with $\Delta t$ ranging from 2 to 40~minutes) and four as \emph{gaps} (with $\Delta t$ from 1.8 to 144.8~hours). 
These weak links and gaps do not imply absent attack evidence: IOC-based validation is independent from reconstruction and confirms the presence of expected artifacts; Recon instead measures the difficulty of recovering full chains from generic telemetry correlation.

While temporal-window-based event correlation is a well-studied primitive in provenance tracking~\cite{cai2026building,10646725} and intrusion detection~\cite{10.1145/3539605}, our contribution lies in combining multiple heterogeneous correlation signals (entity, flow, IP-pair) into a unified graph with a three-level chain assessment---an approach tailored to the multi-host, multi-step scenarios in our dataset.

\paragraph{(4) Metrics, ambiguity, and failure characterization.}
When expected steps are available, we report step- and chain-level precision/recall against the expected step set; otherwise we report coverage-oriented observability proxies. A continuity proxy flags step transitions with excessive temporal gaps (default: 10 minutes). We characterize failures via missing-evidence patterns and, using tagger diagnostics, distinguish true evidence absence from schema/rule mismatches (e.g., missing fields, prefilter drops, or unusable rules). 
We quantify ambiguity at two levels: (i) \emph{event-level} ambiguity as the fraction of events that match multiple step tags ($|\mathcal{M}(e)|>1$), and (ii) \emph{chain-level} ambiguity as competition among candidate chains, measured by the top-2 score margin (and optionally the entropy over top-$K$ candidates).

\paragraph{(5) Source-budgeted runs (Q1--Q3).}
We rerun the same pipeline under controlled source budgets: \emph{single-source} (one stream), \emph{combo-source} (small fixed pairs), and \emph{multi-source} (all available). Per scenario and budget category, we report one representative run selected by reconstruction quality to enable clean cross-scenario comparisons.
All parsers, rules, and reconstruction parameters are fixed across scenarios; differences therefore reflect telemetry availability and attack structure rather than scenario-specific tuning.

\subsection{Scenario Results Overview}

\subsubsection{Collected telemetry summary}

\begin{table}[t]
\centering
\small
\caption{Statistics of Collected Telemetry}
\label{tab:collected_data_stats}
\begin{tabular}{@{}lrrr@{}}
\toprule
\textbf{Scenario} & \textbf{Total Records} & \textbf{Benign Records} & \textbf{Attack Records} \\
\midrule
1.Stegano   & 23{,}534  & 22{,}878  & 656 (2.79\%) \\
2.Starter   & 53{,}978  & 53{,}900  & 78 (0.14\%) \\
3.Parallel  & 88{,}674  & 88{,}385  & 289 (0.33\%) \\
4.NPMEX     & 188{,}270 & 188{,}219 & 51 (0.03\%) \\
5.3CX       & 7{,}453   & 6{,}620   & 833 (11.18\%) \\
6.CloudEX   & 9{,}774   & 9{,}397   & 377 (3.86\%) \\
7.LayerInj  & 222{,}046 & 221{,}411 & 635 (0.29\%) \\
\bottomrule
\end{tabular}
\end{table}

\begin{table}[t]
\centering
\caption{Phase-level IOC observability across all seven scenarios.
Unobserved steps are caused by in-memory C2 execution (9 steps)
or encrypted payload content (1 step), rather than telemetry loss.}
\label{tab:ioc-coverage}
\small
\setlength{\tabcolsep}{4pt}
\begin{tabular}{@{}llcrc@{}}
\toprule
\textbf{Category} & \textbf{ATT\&CK Phase} & \textbf{Observed} & \textbf{Total} & \textbf{\%} \\
\midrule
\multirow{6}{*}{\rotatebox{90}{\scriptsize Infrastructure}}
  & Initial Access    & 8  & 8  & 100 \\
  & Execution         & 8  & 8  & 100 \\
  & C2 Establishment  & 6  & 6  & 100 \\
  & Persistence       & 2  & 2  & 100 \\
  & Defense Evasion   & 1  & 1  & 100 \\
  & Exfiltration      & 6  & 6  & 100 \\
\cmidrule(l){2-5}
  & \textit{Subtotal} & \textit{31} & \textit{31} & \textit{100} \\
\midrule
\multirow{3}{*}{\rotatebox{90}{\scriptsize Post-expl.}}
  & Discovery         & 2  & 6  &  33 \\
  & Collection        & 4  & 9  &  44 \\
  & Credential Access & 0  & 1  &   0 \\
\cmidrule(l){2-5}
  & \textit{Subtotal} & \textit{6}  & \textit{16} & \textit{38} \\
\midrule
\multicolumn{2}{@{}l}{\textbf{Overall}} & \textbf{37} & \textbf{47} & \textbf{78.7} \\
\bottomrule
\end{tabular}
\end{table}

Table~\ref{tab:collected_data_stats} summarizes the volume of normalized telemetry across all scenarios. 
A \emph{record} denotes a single normalized telemetry entry (i.e., one atomic observation) produced from any evidence channel. 

\textbf{Attack records} are defined strictly based on \emph{IOC-level matching}. 
Specifically, for each scenario, we derive a set of expected IOCs from the simulated attack steps and their execution timeline, and manually validate their presence in the collected telemetry. 
A record is labeled as an attack record if it contains at least one such IOC.

To avoid overcounting, we apply a \emph{deduplication} policy at the record level: if a single record matches multiple IOCs, it is counted only once as a single attack record.
Importantly, this labeling is \emph{purely artifact-driven}. 
It does \textbf{not} consider whether the matched record corresponds to a semantically meaningful attack step, nor does it account for triggered side effects such as process creation, subprocess execution, or causal propagation across events. 
As a result, attack records should be interpreted as \emph{validated IOC hits} rather than confirmed execution of attack logic.

Table~\ref{tab:collected_data_stats} characterizes the dataset at the record level, but but record counts alone do not indicate whether the
corresponding attack phases are observable. We therefore add a phase-level IOC observability view in Table~\ref{tab:ioc-coverage} with detailed discussion in Appendix~\ref{app:ioc coverage and data quality}. The results show that infrastructure-facing stages are consistently observable, while blind spots concentrate in post-exploitation behavior caused by in-memory C2 execution or encrypted payload content. This distinction helps interpret the reconstruction results: weak links and gaps reflect limits in correlating phase evidence into continuous chains, rather than simply missing IOC-level telemetry.

\subsubsection{Representative baselines and source-budget comparison.}

\paragraph{Source-budget definition.}
To contextualize our results against common prior settings and to isolate the benefit of additional telemetry, we group configurations by \emph{source budget}---the number of distinct evidence sources available to the pipeline. For each scenario, the corresponding full-telemetry source set (``Multi (full telemetry)'') is enumerated in Appendix~\ref{app:scenario result} (Table~\ref{tab:scenario_budget_breakdown}).
Importantly, \texttt{events} (our exported \texttt{azure\_events} dataset) is a \emph{composite} stream rather than a single-channel log, so configurations that include \texttt{events} may exceed a 2-source budget despite appearing as one dataset (details in Appendix~\ref{app:scenario result}).

\paragraph{Representative configurations.}
\emph{Single-source (1)} baselines operate on one telemetry stream, reflecting host-only provenance/audit or single-stream detectors commonly assumed in prior work (e.g., audit/provenance)~\cite{ProGrapher2023,NODLINK2024,Cheng2023KairosPI}.
\emph{Combo (2)} baselines use exactly two sources; we instantiate a representative \textbf{host+network} setting (\texttt{audit+Zeek}) that combines host causality with network connectivity signals~\cite{APTMCL2026,LOTLHunter2026}.
\emph{Multi ($\ge$3)} settings use three or more sources; as a practical example that frequently appears in deployments and prior work, \textbf{system+events} (\texttt{syslog+events}) falls into this category under our composite-stream accounting~\cite{MultiSourceLogSemantic2025}.
Within \emph{multi-source} ($\ge$3), we additionally evaluate a \emph{full-telemetry} setting that uses the maximum telemetry available in each scenario, representing the strongest achievable configuration of our pipeline.

\paragraph{Metrics rationale.}
We use three metric families to disentangle what is \emph{observable} from what is \emph{correctly reconstructed}. \textbf{Coverage} (TagCov, ChainCov) measures whether the available telemetry exposes the expected coarse steps at all. \textbf{Precision/Recall} (StepP/R, ChainP/R) quantify reconstruction correctness against ground-truth expected steps, where recall captures completeness and precision captures the absence of spurious step attributions. \textbf{Reconstructability} combines recall with a temporal continuity proxy that penalizes excessive gaps between consecutive chain steps. For cross-scenario aggregation, coverage and recall are weighted by the number of expected steps $E_s$ to avoid over-emphasizing simpler scenarios, while precision and reconstructability use unweighted means. Formal definitions are provided in Appendix~\ref{app:metrics}.

\textbf{Q1--Q3 summary.}
We quantify how reconstruction quality changes with increasing source budget.
Table~\ref{tab:source-budget-summary} reports cross-scenario aggregates by budget, and Appendix~\ref{app:scenario result} (Table~\ref{tab:scenario_budget_breakdown}) provides per-scenario best-achievable configurations and missing steps.
We use these summaries to answer Q1--Q3 in terms of observability (Tag/Chain Coverage), detection quality (Step/Chain Recall and Precision), and end-to-end chain quality (Reconstructability).

\begin{table*}[htbp]
\centering
\caption{Cross-scenario summary by source budget and representative combinations. Coverage/recall are weighted by expected steps (wtd.). \textbf{Bold} indicates the best value and \underline{underline} indicates the second-best value in each metric column.}
\label{tab:source-budget-summary}
\small
\setlength{\tabcolsep}{2.8pt}
\renewcommand{\arraystretch}{1.15}
\begin{tabular}{p{4.6cm}rrrrrrrr}
\toprule
Category & \makecell[r]{\{n\}SC} &
\makecell[r]{Tag Cov.\\(wtd.)} &
\makecell[r]{Chain Cov.\\(wtd.)} &
\makecell[r]{StepR\\(wtd.)} &
\makecell[r]{ChainR\\(wtd.)} &
\makecell[r]{StepP\\(mean)} &
\makecell[r]{ChainP\\(mean)} &
\makecell[r]{Recon.\\(mean)} \\
\midrule
Single (1): avg over all single sources
& 6 & 0.263 & 0.263 & 0.263 & 0.263 & \underline{0.775} & \underline{0.775} & 0.266 \\
Single (1): best single-source
& 6 & 0.391 & 0.391 & 0.391 & 0.391 & \textbf{1.000} & \textbf{1.000} & 0.403 \\
Single (1): audit/provenance \cite{ProGrapher2023,NODLINK2024}
& 3 & 0.091 & 0.091 & 0.091 & 0.091 & 0.333 & 0.333 & 0.111 \\
Single (1): Zeek \cite{ZeekBeaconing2025,CobaltStrikeMetadata2025}
& 3 & 0.273 & 0.273 & 0.273 & 0.273 & \textbf{1.000} & \textbf{1.000} & 0.278 \\
\midrule
Combo (2): avg over 2-source pair
& 5 & 0.430 & 0.400 & 0.430 & 0.400 & \textbf{1.000} & \textbf{1.000} & 0.431 \\
Combo (2): best 2-source pair
& 3 & \textbf{0.636} & \textbf{0.545} & \textbf{0.636} & \textbf{0.545} & \textbf{1.000} & \textbf{1.000} & \textbf{0.639} \\
Combo (2): audit+Zeek \cite{APTMCL2026,LOTLHunter2026}
& 3 & 0.364 & 0.273 & 0.364 & 0.273 & \textbf{1.000} & \textbf{1.000} & 0.389 \\
\midrule
Multi ($\ge$3): syslog+events \cite{MultiSourceLogSemantic2025}
& 2 & \underline{0.500} & \underline{0.500} & \underline{0.500} & \underline{0.500} & \textbf{1.000} & \textbf{1.000} & \underline{0.500} \\
Multi ($\ge$3): avg full telemetry
& 7 & 0.481 & 0.481 & 0.481 & 0.481 & \textbf{1.000} & \textbf{1.000} & 0.488 \\
Multi ($\ge$3): best full telemetry
& 7 & 0.481 & 0.481 & 0.481 & 0.481 & \textbf{1.000} & \textbf{1.000} & 0.488 \\
\bottomrule
\end{tabular}

\vspace{2pt}
\footnotesize{\textbf{Note}: events data is taken as a composite telemetry stream (multiple evidence channels).}
\end{table*}

\paragraph{Q1: How completely can end-to-end attack chains be reconstructed from the available telemetry?}
Under full telemetry, our pipeline reaches $\mathrm{TagCov}^{\mathrm{wtd}}=0.481$ and $\mathrm{StepR}^{\mathrm{wtd}}=0.481$ across all seven scenarios, with mean reconstructability $0.488$. This indicates that, on average, multi-source evidence substantially improves end-to-end reconstruction relative to single-source settings (best single: $\mathrm{TagCov}^{\mathrm{wtd}}=0.391$, $\mathrm{StepR}^{\mathrm{wtd}}=0.391$, reconstructability $0.403$). At the scenario level (Table~\ref{tab:scenario_budget_breakdown} in Appendix~\ref{app:scenario result}), reconstruction remains bimodal: SC2 and SC4 achieve $\mathrm{StepR}=0.75$ (missing only \texttt{OUTBOUND\_CONN} and \texttt{EXFIL} respectively), while SC1/SC5/SC6 remain at $\mathrm{StepR}=0.25$ due to persistent absence of \texttt{DOWNLOAD}/ \texttt{\seqsplit{OUTBOUND\_CONN}}/ \texttt{EXFIL} evidence. SC7 reaches $\mathrm{StepR}=0.667$ but consistently misses \texttt{DOWNLOAD}, consistent with model-level attacks whose retrieval phase is weakly expressed in available host/network schemas. Notably, precision remains 1.000 under all multi-source configurations, indicating that the pipeline does not misattribute benign activity as attack steps.

\paragraph{Q2: Where and why does single-source telemetry fail?}
Single-source telemetry exhibits two systematic failure modes. First, evidence incompleteness: averaged over all single sources, $\mathrm{TagCov}^{\mathrm{wtd}}$ and $\mathrm{StepR}^{\mathrm{wtd}}$ both drop to 0.263, with mean precision 0.775. The gap between ``avg over all singles'' and ``best single'' reflects that many single-source runs observe few of the expected steps, yielding undefined precision that we conservatively treat as zero. Second, semantic/causal ambiguity: even the best-achievable single-source selection (best single) systematically misses cross-layer phases such as \texttt{DOWNLOAD} and \texttt{EXFIL} (e.g., SC2 and SC4 in the Appendix~\ref{app:scenario result}), because these steps require joinable host execution context and network/service evidence that a single stream cannot provide.

\paragraph{Q3: How does multi-source mitigate these failures, and what remains unresolved?}
Multi-source mitigates single-source failures primarily by improving completeness: adding complementary anchors increases observability and raises aggregate recall from 0.391 (best single) to 0.481 (full telemetry), while improving mean recon from 0.403 to 0.488. The gains are driven mainly by scenarios where missing phases become observable under additional sources (e.g., SC2 and SC3/SC4 in the Appendix~\ref{app:scenario result}). However, multi-source does not universally resolve missing-step gaps: SC1/SC5/SC6 remain bounded under our conservative rule-based anchors, as manual audit shows that some phases are not captured as portable, reproducible chain anchors or remain semantically ambiguous; additional sources help only when they expose the missing phase with compatible join keys, rather than merely adding log volume. Finally, two-source pairs can outperform or match multi-source on the subset of scenarios where such pairs exist and are highly informative (Combo best: reconstructability 0.639 over 3 scenarios in Table~\ref{tab:source-budget-summary}); nevertheless, the primary benefit of multi-source is \emph{robustness across heterogeneous scenarios}, not dominance on every scenario subset.

\paragraph{Validation checks.}
We add two validation checks for our rule-based reconstruction. 
First, using only defender-visible raw telemetry, we audit all 14 missing \texttt{DOWNLOAD}, \texttt{OUTBOUND\_CONN}, and \texttt{EXFIL} anchors reported in Appendix~\ref{app:scenario result}: 10 are conservative-rule misses and four are semantic ambiguities, with none requiring attacker-side evidence (Appendix~\ref{app:csa3_details}). Thus, the reconstruction metrics are reproducible lower bounds on rule-matchable anchors rather than exhaustive measures of raw-telemetry evidence.
Second, although \texttt{INSTALL/DOWNLOAD} signals overlap with benign package activity, adding downstream chain context reduces benign candidate windows from 74 to 21 (71.6\%) while retaining all seven scenarios. This candidate-volume result indicates that chain reconstruction can support detection triage in addition to post-hoc forensics, helping defenders assess attack progression and prioritize containment (Appendix~\ref{app:csa3_details}).

\subsection{Case Studies}

We present two contrasting scenarios to illustrate both the strengths and the evidence-bound limits of our telemetry-to-chain pipeline; full evidence packages are in Appendix~\ref{app:evidence}.

\subsubsection{SC4: NPMEX --- Sequential Dependency Chain Attack (Positive exemplar: near-complete chain)}
\label{sec:sc4_npmex}

The full-telemetry configuration reconstructs three of the four expected coarse steps, achieving $\mathrm{StepR}=0.75$ with perfect step precision ($\mathrm{StepP}=1.0$).
The observed step set is \{\texttt{INSTALL}, \texttt{DOWNLOAD}, \texttt{OUTBOUND\_CONN}\} and the reconstructed chain is \texttt{\seqsplit{OUTBOUND\_CONN}} $\rightarrow$ \texttt{INSTALL} $\rightarrow$ \texttt{DOWNLOAD}.
Across the run, the pipeline ingests 188{,}270 normalized records; the strongest evidence comes from high-volume network telemetry for \texttt{\url{OUTBOUND\_CONN}} (connection records), complemented by a small number of high-specificity host records for \texttt{INSTALL} (package-manage actions) and \texttt{DOWNLOAD} (explicit retrieval commands).

\paragraph{Analysis}
SC4 is reconstructable because it provides \emph{complementary anchors with compatible join keys}.
Network telemetry (e.g., Suricata/Zeek) supplies stable connection-level evidence that grounds \texttt{\seqsplit{OUTBOUND\_CONN}} in time and endpoints, while host telemetry (syslog/auth) provides execution-context anchors for \texttt{INSTALL} and explicit fetch behavior for \texttt{DOWNLOAD}.
These anchors are temporally consistent and share joinable entities (host identity, process/user context, and/or endpoints), allowing the event graph to connect phases into a coherent chain.

\noindent\textbf{Why no \texttt{EXFIL}}
We do not instantiate a separate \texttt{EXFIL} node because the observed package logic is consistent with a loader--executor design: it performs token bootstrap and payload retrieval/execution, but does not implement an explicit ``collect $\rightarrow$ serialize $\rightarrow$ send'' routine in the published artifacts. Consequently, any exfiltration would be attributable only to the downloaded second-stage script, and lacks a distinctive host-side marker that would support a reliable, joinable \texttt{EXFIL} anchor in our reconstruction.

\subsubsection{SC1: Stegano — Steganography Exploitation (Negative exemplar: evidence-bound ceiling)}
\label{sec:sc1_stegano}

\begin{figure}[htbp]
\centering
\includegraphics[width=\columnwidth]{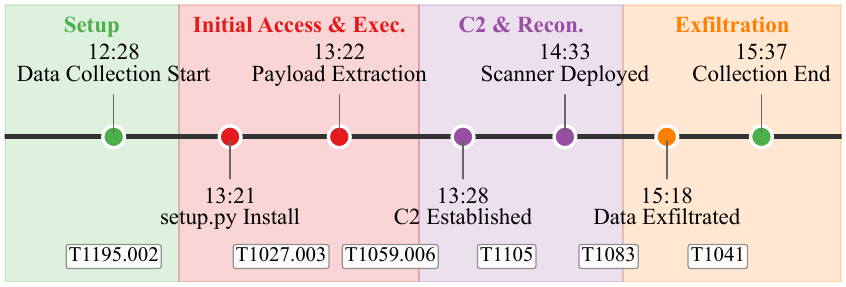}
\caption{SC1 attack lifecycle}
\label{fig:sc1_timeline}
\end{figure}

The attack unfolds across five phases, visualized in Figure~\ref{fig:sc1_timeline} with corresponding MITRE ATT\&CK technique identifiers at each transition. Reconstruction is bounded despite full telemetry ingestion. The full-telemetry run observes only \texttt{INSTALL} from the expected set \{\texttt{INSTALL}, \texttt{DOWNLOAD}, \texttt{\seqsplit{OUTBOUND\_CONN}}, \texttt{EXFIL}\}, yielding $\mathrm{StepR}=0.25$ (with $\mathrm{StepP}=1.0$).
Although the run ingests 8{,}534 normalized records, step-tagging fires only sparsely and concentrates on installation-related process activity (e.g., the package installation command), while no rule-matchable evidence is produced for \texttt{DOWNLOAD}, \texttt{\seqsplit{OUTBOUND\_CONN}}, or \texttt{EXFIL}.

\paragraph{Analysis}

SC1 illustrates an \emph{evidence-bound failure mode}: adding more telemetry sources increases volume but does not necessarily increase \emph{usable} anchors.
Here, the expected phases \texttt{DOWNLOAD}/ \texttt{\url{OUTBOUND\_CONN}}/ \texttt{EXFIL} are not recovered as portable rule-matchable anchors: some evidence is only indirectly visible, lacks stable join keys, or requires semantic interpretation beyond our conservative tagger.
As a result, the event graph lacks the anchors needed to connect installation to subsequent phases, and multi-source correlation cannot compensate for missing or non-joinable evidence.
This negative case motivates our failure taxonomy and deployment implications: multi-source telemetry helps only when it reveals the missing phase with joinable entities (process/user/endpoint), rather than merely adding additional records.

\section{Cross-Scenario Analysis}
\label{sec:cross-scenario analysis}

To interpret the aggregate results in Table~\ref{tab:source-budget-summary}, we analyze cross-scenario mechanisms that govern reconstruction quality and ATT\&CK coverage. The goal is to identify which properties transfer across scenarios (stable anchors and join keys) versus which are scenario-structural (missing phases, cloud/control-plane actions), and to distill actionable guidance for telemetry planning. 

\subsection{CSA-1: Attack Chain Reconstructability}

Reconstructability is governed primarily by (i) whether each phase exposes a stable telemetry anchor and (ii) whether anchors share joinable identifiers across sources (host/user/process and network endpoints). When these conditions hold, causal chaining is reliable; otherwise the pipeline conservatively outputs partial chains rather than brittle full narratives (Appendix~\ref{app:csa1_details} ).

\subsection{CSA-2: TTP Observability and Alignment}

We align reconstructed coarse steps to ATT\&CK post hoc, because the same attacker action can project differently across telemetry layers. Alignment is therefore evidence-conditioned: missing projections or weak attribution can under-support techniques even when the attack occurred. Scenario-level ATT\&CK breadth and time window further modulate alignment difficulty (Appendix~\ref{app:csa2_details}).

\subsection{CSA-3: Failure Taxonomy \& Observability}

Across scenarios, failures fall into three dominant classes: missing-phase evidence (true observability gaps), non-joinable evidence (attribution breaks), and ambiguity/noise (generic events or concurrency). Multi-source helps mainly when it adds the missing phase or strengthens joins; it cannot recover phases that remain absent, non-joinable, or only semantically inferable under our conservative rule-based anchors (see Table~\ref{tab:failure_minimal} in Appendix~\ref{app:csa3_details} for details).

\subsection{CSA-4: Sensitivity to Telemetry Density}
\label{sec:csa4-sensitivity}

In practice, telemetry is often subject to sampling, or partial collection due to cost and performance constraints~\cite{dempsey2011iscm}. To assess robustness, we conduct a preliminary sensitivity analysis by varying the per-source \emph{sampling rate} and measuring how reconstruction performance degrades across source-budget settings. For each telemetry source, we retain each raw event independently with a series of sampling rates (Bernoulli downsampling~\cite{dempsey2011iscm}), applied \emph{before} step tagging to faithfully model reduced evidence availability. We repeat each configuration with 5 random seeds and report mean $\pm$ standard deviation across seeds.

\begin{figure}[t]
  \centering
  \includegraphics[width=\columnwidth]{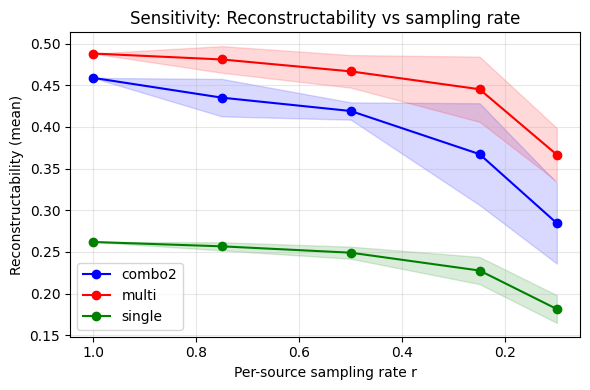}
  \caption{Sensitivity of recon to per-source sampling rate.}
  \label{fig:sens-recon}
\end{figure}

Figure~\ref{fig:sens-recon} plots mean recon against sampling rate for each budget category. Multi-source configurations exhibit a shallower degradation curve: at $r{=}0.25$, multi-source retains a higher fraction of its baseline recon compared to single-source, confirming that source diversity provides natural robustness to evidence sparsity.

This is consistent with CSA-3: when one stream loses events due to sampling, complementary anchors from other streams can compensate, provided joinable identifiers remain intact.
Conversely, single-source performance degrades more steeply, as evidence loss in a single stream directly eliminates anchors with no fallback.
The two-source setting falls in between, with the gap depending on whether the paired sources cover complementary phases.
Full per-metric breakdowns are provided in Appendix~\ref{app:csa4_full_sensitivity}.

\subsection{CSA-5: Patterns \& Deployment Implications}

Two-source host+network budgets can be strong when phases are joinable, but multi-source is most valuable for robustness across heterogeneous attack structures (cloud control-plane actions, enterprise software, and model-layer attacks). Telemetry planning should prioritize diverse evidence types and stable join keys over log volume; targeted additions (e.g., IAM/API audit streams) yield outsized gains for structurally hard cases (Appendix~\ref{app:csa5_deployment_implications}).

Importantly, these guidelines do not rely on prior knowledge of specific attacks. Instead, they operationalize telemetry selection based on structural observability requirements. In practice, defenders can: (i) ensure coverage of fundamental attack phases (e.g., ingress, execution, outbound communication, exfiltration),
(ii))prioritize sources that provide stable cross-layer join keys, and (iii) incrementally diagnose blind spots using partial detections (e.g., unmatched or truncated chains).
This allows defenders to iteratively refine telemetry configurations without requiring ground-truth knowledge of ongoing attacks, using reconstruction failures themselves as signals of missing visibility.

%% file: sections/discussion.tex
\section{Discussion}
\label{sec:discussion}

\paragraph{Scope and constraints.}
SynthChain targets the exploitation stage of SSC attacks on an end-user host, focusing on the observable behaviors after a malicious artifact is introduced and executed. As a result, we do not aim to model the full lifecycle of long-running APT campaigns, such as large-scale lateral movement or complex privilege escalation, which are already well represented in existing datasets \cite{MYNENI2023109688,edq8-nk52-21}. This scope is consistent with the operational profile of SSC exploitation: because attacks leverage \textbf{default trust} in package managers and build pipelines, the chain from installation trigger to exfiltration typically completes within a single session, as observed in the incidents~\cite{hiddenplainsight,lazarusgroup,3CX}. Moreover, our primary goal is to \textbf{measure observability limits} — whether each attack phase produces joinable evidence in available telemetry — which is a structural property independent of attack duration.

While our evaluation is scenario-driven rather than Internet-scale, it is telemetry-rich: we collect large-volume multi-source traces under controlled source budgets to support systematic cross-layer reconstruction.
We exclude macOS and all BSD-variants: compared to Linux and Windows, its stronger built-in protections restrict audit visibility, limiting telemetry availability and comparability~\cite{apple_es_api}. We therefore focus on Linux and Windows for richer, more consistent security auditing and multi-source alignment.

\paragraph{Rule granularity and matching assumptions.}
Step tagging and part of the reconstruction rely on coarse, portable rule matching (e.g., regular expressions and keywords) over normalized fields. This choice improves cross-scenario comparability, but it may under-approximate behaviors when telemetry schemas differ or when benign software shares similar surface tokens. Steps such as \texttt{INSTALL, DOWNLOAD} and dependency/lifecycle-hook actions are particularly difficult to disambiguate from textual fields alone, which can reduce precision. 

\paragraph{Future work: topology-aware provenance for software and dependencies.}
This direction aims to achieve higher-fidelity alignment to scenario ground truth (where available), reducing ambiguity beyond coarse token-based matching rather than claiming exact one-to-one ground-truth matches.
To move beyond token-level matching, we plan to incorporate topology-aware evidence extraction \cite{10120960, 10.5555/3766078.3766446}. Concretely, we plan to incorporate (i) process and file provenance graphs (parent--child execution and file write--read chains), (ii) package-manager and dependency resolution traces, and (iii) network-to-process attribution using richer host tracing (e.g., eBPF) to recover higher-fidelity causal links among download, installation, and subsequent execution. Such topology-driven correlation would reduce reliance on surface tokens and provide more stable join structures across heterogeneous telemetry schemas, improving accuracy and robustness of chain reconstruction.

%% file: sections/conclusion.tex
\section{Conclusion}
\label{sec:conclusion}

SSC attacks are inherently multi-stage and cross-layer, so their evidence is fragmented across heterogeneous telemetry streams and often lacks stable anchors for cross-source alignment, making end-to-end reconstruction challenging under common single-source assumptions.
To address this gap, we introduce SynthChain, a SSC-centric dataset with diverse multi-stage scenarios, multi-source telemetry collection, 2{,}919 manually verified IOC annotations across 22 log files spanning 11 telemetry types, and explicit cross-source alignment, enabling systematic evaluation of telemetry-to-chain reconstruction under controlled source budgets.

Our cross-scenario analysis yields three transferrable insights: (i) reconstruction depends on whether each phase exposes a stable telemetry anchor and whether anchors share identifiers across sources; (ii) single-source telemetry fails due to missing phases and semantic/causal ambiguity, often dropping or misattributing steps; and (iii) multi-source correlation improves completeness via complementary anchors and joins, but cannot recover phases that are absent or non-joinable and may add spurious candidates in ambiguous settings (e.g., model-layer attacks). A sensitivity analysis further confirms that the multi-source advantage persists under reduced per-source sampling rates, suggesting that complementary source selection remains effective even under realistic evidence sparsity constraints. The open-source release of SynthChain can catalyse a step-change in empirical research by making multi-stage, multi-source reconstruction a shared, reproducible benchmarking challenge for complex SSC attacks.